\documentclass[pre,aps,floats,superscriptaddress,floatfix,twocolumn]{revtex4}
\usepackage{amssymb,amsmath}
\usepackage{amsmath,amssymb}
\usepackage{graphicx}
\usepackage{psfrag}
\usepackage{color}
\usepackage{soul}
\usepackage{dcolumn}
\usepackage{bm}
\usepackage[normalem]{ulem}
\newcommand{\cred}{\color{red}}

\def\beq{\begin{equation}}
\def\eeq{\end{equation}}
\def\bea{\begin{eqnarray}}
\def\eea{\end{eqnarray}}
\begin{document}
 \title{Nonuniform asymmetric exclusion process: Stationary densities and domain walls}
 \author{Sudip Mukherjee}\email{sudip.bat@gmail.com,aca.sudip@gmail.com}
\affiliation{Barasat Government College,
10, KNC Road, Gupta Colony, Barasat, Kolkata 700124,
West Bengal, India}
\author{Abhik Basu}\email{abhik.123@gmail.com, abhik.basu@saha.ac.in}
\affiliation{Theoretical Physics Division, Saha Institute of
Nuclear Physics, 1/AF Bidhannagar, Calcutta 700064, West Bengal, India}
\date{\today}
\begin{abstract}
 We compute the stationary densities in totally asymmetric exclusion processes (TASEP)
with open boundary conditions and spatially nonuniform hopping rates. 
The stationary densities in the low and and high density phases can be discontinuous, only when the space-dependent hopping rate is spatially discontinuous.
In contrast, the stationary density profile in the maximal current phase can be discontinuous, even when the space-dependent hopping rate is continuous. We further investigate the domain walls, which are delocalised with complex shapes. { In striking contrast to a delocalised domain wall (DDW) in an open, uniform TASEP, these DDWs can also form at the transitions between the low or high density phases and maximal current phase, and can cover the entire TASEP channel or a part of it, depending upon the specific forms of the site-dependence of the hopping rates and the associated phase transitions.} We calculate their envelopes, which are curved lines, revealing their dependence on the spatial nonuniformity of the hopping rates. The phase diagrams in the plane of the control parameters show universal topology. The associated phase transitions are explored, which can be different from their counterparts in an open uniform TASEP.  
\end{abstract}

\maketitle

\section{Introduction}


Totally asymmetric simple exclusion process (TASEP) is one of the best studied nonequilibrium models~\cite{tasep-rev1,tasep-rev2,tasep-rev3}. It was originally proposed as a simple model for protein synthesis in eukaryotic cells~\cite{macdonald}. Later on, it has been reinvented as a paradigmatic one-dimensional (1D) model for nonequilibrium phase transitions in open systems~\cite{tasep-rev1}. A TASEP is defined as a
1D lattice gas, where particles are hopping stochastically in one direction subject to hard-core repulsion or {\em exclusion}; see, e.g., Ref.~\cite{derrida-rev}.
Subsequently, motivated by a variety of phenomena of biological and social origins, many generalisations of TASEP have been proposed and studied. These studies have revealed novel nonequilibrium phenomena and rich phase behaviours. For example, it has been found from a large number of studies that a TASEP lane coupled with weak particle nonconserving processes, e.g., exchange of particles with the embedding medium or with another  lane executing diffusive dynamics or even another, second TASEP lane, or connected to a network, or undergo reactions can display complex nonuniform steady states of various kinds~\cite{erwin-lk-prl,erwin-lk-pre,tobias-ef1,tobias-ef2,tobias-ef3,rakesh1,rakesh2,arvind1,frey-graf-PRL,bojer-graf-frey-PRE,arvind2,sm,sourav5,atri4}. Similarly, the coupled dynamics of a TASEP lane diffusively exchanging particles with an embedding three-dimensional medium has been explored in Refs.~\cite{klump-PRL,klump-JSTAT,ciandrini-2019,dauloudet-2019}, revealing the nontrivial effects of three-dimensional diffusion on the stationary states in the TASEP lane. 

A particularly interesting generalisation of TASEP involves a TASEP lane, open or closed with quenched disorder or defects. Two distinct kinds of disorders have been considered - particle-wise disorder, in which hopping rates may  depend on the particle
attempting to jump~\cite{part1,part2,part3}, or site-wise disorder, in which each site is associated with a quenched disordered hopping rate~\cite{lebo,mustansir,niladri-tasep,tirtha-niladri,tirtha-lk1,tirtha-lk2,astik-prr,atri3}.  Such site-wise disordered TASEP can be applied to wide-ranging phenomenologies. This includes, for instance, protein synthesis by messenger RNA in the presence of pause sites~\cite{tom-prl,beate1,krug-new,bao-natcomm,mrna5}. For example, FMRP (a polyribosome-associated neuronal RNA-
binding protein) can stall ribosome translocation, or, the
slowdown of a translating ribosome upon encountering a
rare ``slow codon'' in the mRNA due to the low concentration
of the corresponding tRNA~\cite{mrna4}. These ``pause sites'' are believed to have several biological functions; see, e.g., Refs.~\cite{cod1,cod2}. Yet another application of site-wise disordered TASEP is on vehicular or pedestrian traffic along urban transport network with road blocks or local constrictions that can lead to traffic jams~\cite{comp,ped2,ped3}.
Lack of overall theoretical understanding of how quenched disorders may affect the dynamics and
steady states of nonequilibrium models~\cite{noneq1} and the absence of any general framework to study this question have prompted scientists to study the effects of quenched disorder on TASEP as a minimal conceptual model.

In this work, we focus on a specific sub-class of a TASEP with quenched disorder hopping rates, in which the hopping rate varies {\em smoothly} (as opposed to randomly), with position along the TASEP lane~\cite{stinch1,laka,stinch2,tirtha-prr,erdmann,atri-jstat}. In Ref.~\cite{tirtha-prr}, a TASEP in a ring geometry with smoothly varying hopping rate with  finite number of discontinuities has been studied by mean-field theory (MFT) and Monte-Carlo simulations (MCS), revealing generic nonuniform steady states and domain walls. Analogous study for an open TASEP with smoothly varying hopping rates with one minimum can be found in Ref.~\cite{atri-jstat}, which reveals delocalised domain walls (DDW) of complex shapes and phase transitions marked by discontinuous changes in the average bulk densities between different phases. In this work, we generalise the studies in Ref.~\cite{atri-jstat} by choosing spatially slowly varying hopping rates that either have isolated finite discontinuities or have more than one global minimum. 
We use MFT and MCS for our studies.  Our principal results are as follows: (i) We identify the technical difficulties in applying MFT na\"ively to calculate the stationary densities and then demonstrate how these can be remedied by employing MFT together with physically intuitive arguments to analytically construct the correct stationary density profiles, which agree with the MCS results. { This is particularly relevant in applying MFT to the maximal current (MC) phase. (ii) Our intuitive arguments, supported by detailed  MFT results, show that the phase diagrams generally have the same topology as the one for a uniform open TASEP, and may even be {\em exactly} identical to that for an open uniform TASEP, which can be achieved by tuning the space-dependence of the hopping rate. This rules out phase diagrams as reliable markers for different choices of the spatially nonuniform hopping rates. Intriguingly, the transitions between the low density (LD) and maximal current (MC) phases, and those between high density (HD) and MC phases can be discontinuous or continuous transitions, in contrast to uniform open TASEPs, where these two transitions are always continuous transitions~\cite{tasep-rev1,tasep-rev2,tasep-rev3}.  Interestingly, we show that even when the nature of the LD-MC or HD-MC transition is continuous, i.e., same as the corresponding transition in a uniform open TASEP, the MFT gives that the associated order parameter critical exponent describing the transition is {\em different} from its pure TASEP counterpart.  (iii) By systematically going beyond the MFT, we analytically construct the envelops of the DDWs for some representative examples of spatially nonuniform hopping rates. These studies generalise a DDW in the conventional open TASEP with a uniform hopping rate. Since the LD-MC or HD-MC transitions can also be discontinuous with spatially nouniform hopping rates, DDWs can be formed at the these transitions as well (only when the corresponding transition is discontinuous). This is in stark contrast with an open uniform TASEP, where DDWs can never form at the the LD-MC or HD-MC transitions (since there are always continuous transitions in open uniform TASEPs). More intriguingly and being dependent on the precise form of space dependence of the hopping rates, these DDWs at the LD-MC and HD-MC transitions, can be {\em full}, i.e., fully covering the TASEP lane, or {\em partial}, i.e., partially covering the TASEP lane, a unique feature of nonuniform open TASEPs. In addition, DDWs always form at the LD-HD transitions (just like for open uniform TASEPs), which are always {\em full}. For some specific choices of the hopping rate functions, the nature of the long-time averaged density at the meeting point of the LD-HD, LD-MC and HD-MC phase boundaries in the plane of the control parameters can have a rather unusual structure, which may be interpreted as two partial DDWs joined serially at a point in the bulk of the system, which is also the location of the inimum of the hopping rate function.  In general, the DDW envelops can have complex shapes, which are in general nonuniversal and controlled by the precise nature of the space dependence of the hopping rates.} We demonstrate the robustness of our calculational scheme by calculating the DDW envelops for site-dependent hopping rates with finite discontinuities, which agree well with the corresponding MCS results. (iii)  Finally we study the phase diagrams and associated phase transitions. We establish that phase diagrams generally have the same topology as the one for a uniform open TASEP, and may even be {\em exactly} identical to that for an open uniform TASEP, which can be achieved by tuning the space-dependence of the hopping rate. This rules out phase diagrams as reliable markers for different choices of the spatially nonuniform hopping rates. Intriguingly, we show that even when the nature of the transition is continuous and same as the corresponding transition in a uniform open TASEP, the MFT gives that the associated order parameter critical exponent describing the transition is {\em different} from its pure TASEP counterpart.  A brief account of the present study is available in Ref.~\cite{sm-ab-arxiv}; see also Ref.~\cite{thesis} for similar investigations in this context.
The remainder of this article is organised as follows. We lay out the construction of our model in Section~\ref{model}. In Section~\ref{st-den}, we study the stationary densities and construct the MFT for our model for a few representative examples of the space-dependent hopping rates. Then in Section~\ref{ddw} we construct the DDW profiles for various different space dependent hopping rates. Next in Section~\ref{phase}, we obtain the phase diagrams and study the associated phase transitions. Then in Section~\ref{monte}, we briefly discuss our simulation algorithm. Finally, we summarise and discuss our results in Section~\ref{summ}.


\section{Model}\label{model}

\begin{figure}[htb]
 \includegraphics[width=1\columnwidth]{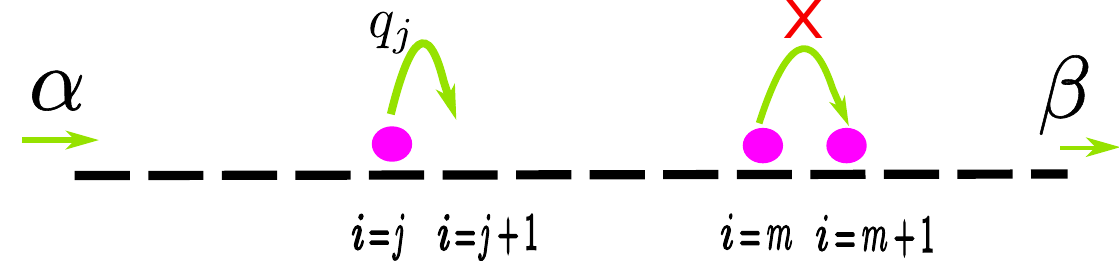}
  \caption{ Schematic model diagram. The broken line represents the TASEP lattice. Particles enter and exit at rates defined by $\alpha$ and $\beta$, respectively, and hop from left to right in the bulk, subject to exclusion. Hopping rate $q_i$ at site $i$ depends explicitly on the site index $i$.}\label{modelfig}
 \end{figure}

In this Section, we construct our model.
We consider an inhomogeneous TASEP on  a 1D lattice with $L$ sites, labelled by $i\in [1,L]$. As in the uniform TASEP model~\cite{tasep-rev1,tasep-rev2,tasep-rev3}, particles enter through the left end ($i=1$) at rate parametrised by $\alpha$, hop unidirectionally from the left to the right, all subject to exclusion,  and finally leave the system at site $i=L$ with a rate parametrised by $\beta$; $0\leq \alpha,\beta\leq 1$. The hopping rate at site $i$ depends upon $i$ and is given by $q_i\leq 1$; see Fig.~\ref{modelfig} for a schematic model diagram. The dynamics is subject to exclusion at all sites and every time step.
 
 Hard core repulsion between neighbouring particles is imposed. This prohibits multiple occupancy of sites in the model. The complete state space of our model then consists of $2^L$
configurations. The following elementary processes fully define the microscopic dynamical update rules of this model:

\noindent (a) At any site $i = 1 , . . . , L-1$ a particle can jump to its neighbouring site $i+ 1$ subject to exclusion with a rate $q_i\leq 1$.

\noindent (b) At the site $i = 1$ a particle can enter the lattice with rate $\alpha q(1)$
provided  it is unoccupied; and

\noindent (c) At the site $i = N$ a particle can leave the lattice with rate
$\beta q(L)$, if there is a particle there, i.e., when it is occupied.

In general, $q_i$ is $i$-dependent, i.e., $q_i\neq q_j$ for $i\neq j$ in our model. Dynamical update rules (a)-(c) formally define an inhomogeneous TASEP with open boundary conditions. If all of $q_i=1$ for all $i$ identically, then this model reduces to a pure TASEP with open boundary conditions and unit hopping rate~\cite{tasep-rev1}. In general, the form of the stationary densities should depend upon the specific choices of $q_i$~\cite{atri-jstat}.  We generalise the results of Ref.~\cite{atri-jstat} and consider some specified choices of $q_i$ that depends explicitly on $i$, and study their effects on the nonequilibrium steady states of the model below.

Before we embark on our studies below, we recall that the steady states of an open TASEP with $\alpha$ and $\beta$ as the entry and exit rates, and a uniform hopping rate are characterised by the mean bulk density $\rho_T$: For $\alpha<\beta$ and $\alpha<1/2$, one has $\rho_T=\alpha$ giving the low density (LD) phase, for $\beta<\alpha$ and $\beta<1/2$, one has $\rho_T=1-\beta$ giving the high density (HD) phase, and for $\alpha,\,\beta>1/2$, one has $\rho_T=1/2$ giving the maximal current (MC) phase. This immediately gives $\alpha=\beta,\,0\leq \alpha,\beta\leq 1/2$ as the phase boundary, or the coexistence line between the LD and HD phases in the $\alpha-\beta$ plane~\cite{tasep-rev1}. The density profile for $\alpha,\beta$ on this coexistence line is a DDW~\cite{kolom}. The principal aim of the present study is to find the phases, phase boundaries and the nature of the associated phase transitions, and the principles behind obtaining them when the hopping rate is not constant in space, but spatially smoothly varying. 

\section{Steady state densities}\label{st-den}

Our aim is to study the stationary densities in the steady states and explore the domain walls, which are moving or {\em delocalised}, i.e., DDWs. For this purpose, we first set up an MFT, amenable to analytical solutions for the steady state densities. We show how it can be applied to obtain unambiguous predictions on the steady state density profiles. By going beyond MFT, we further calculate the envelope of the DDWs analytically, which show that the DDW profiles actually depend rather {\em weakly} on the specific natures of space-dependent hopping rates. We supplement our analytical results by extensive MCS studies, which agree well with our analytical solutions.

 \subsection{Mean-field theory}\label{mean}
 
 We first consider the rate equations defined above for every site, which are {\em not} closed. The MFT approximation entails neglecting correlation effects and replacing the average of product of densities by the product of average of densities~\cite{blythe}. Although this is an approximation, this has worked with a good degree of accuracy in the original TASEP problem and also later in  many of its variants (see, e.g., Refs.~\cite{erwin-lk-pre,niladri-tasep,tirtha-niladri}).  The MFT dynamical equation for $n_i(t)$ reads~\cite{atri-jstat}
 \begin{equation}
  \frac{\partial n_i}{\partial t} = q_{i-1}n_{i-1}(1-n_i) - q_{i}n_i(1-n_{i+1}),\label{basiceq-0}
 \end{equation}
  for a site $i$ in the bulk.  Our model  exhibits {\em particle-hole symmetry} in the following sense: a jump of a particle to the
right corresponds to a vacancy or hole movement by one step to the left. Similarly, a particle entering the TASEP channel at the left
boundary can be interpreted as a hole leaving it through
the left boundary, and vice versa for the right boundary. Indeed, Eq.~(\ref{basiceq-0}) is invariant under the transformation $n_i(t) \rightarrow 1-n_{L-i}(t)$ together with $q_i \rightarrow q_{L-i}$ and $\alpha \rightarrow \beta$, which formally defines the {\em particle-hole symmetry} in this model~\cite{erwin-lk-pre,atri-jstat}. In our subsequent analysis below, for the ease of presentation of our results from stochastic MCS studies, we take a continuum limit with $\rho(x)\equiv n_i$, with $x=i/L$ becoming quasi-continuous in the limit $L\rightarrow\infty$ and $0\leq x\leq 1$.  In this parametrisation, the discrete hopping rates $q_i$ are given by $q(x)$, with $0<q(x)\leq 1$.  Thus for every inhomogeneous TASEP characterised by inhomogeneous hopping rates $q_i$, one can construct a unique $q(x)$. In this work, we consider few specific examples of $q(x)$, which  vary {\em slowly} with $x$. We define a lattice constant $\varepsilon \equiv L_0/L$, where $L_0$ is the geometric length of the system, which has been set to unity without any loss of generality. In the limit $L\rightarrow \infty$, $\varepsilon\rightarrow 0$ is a small parameter in the problem. As shown in Refs.~\cite{tirtha-prr,atri-jstat}, the general solution for the stationary density is 
\begin{equation}
 \rho(x)=\frac{1}{2}\left[1\pm \sqrt{1-\frac{4J}{q(x)}}\right],
\end{equation}
where $J$ is the stationary current, a constant throughout the system. In MFT, $J$ depends on the specific phase the TASEP lane is in. We have
 \begin{eqnarray}
 J_\text{LD}&=&q(0)\alpha(1-\alpha),\label{j-ld}\\
 J_\text{HD}&=&q(1)\beta(1-\beta),\label{j-hd}\\
 J_\text{MC}&=& \frac{q_\text{min}}{4} \label{j-mc}
\end{eqnarray}
are, respectively, the currents in the LD, HD and MC phases of the TASEP lane. With these stationary currents defined in (\ref{j-ld})-(\ref{j-mc}), the stationary densities in the LD, HD and MC phases are~\cite{atri-jstat}
\begin{eqnarray}
 \rho_\text{LD}(x)&=&\frac{1}{2}\left[1-\sqrt{1-\frac{4J}{q(x)}}\right]<1/2,\,J=J_\text{LD},\label{rho-ld}\\
 \rho_\text{HD}(x)&=&\frac{1}{2}\left[1+\sqrt{1-\frac{4J}{q(x)}}\right]>1/2,\,J=J_\text{HD},\label{rho-hd}\\
 \rho_\text{MC}(x) &=& \frac{1}{2}\left[1-\sqrt{1-\frac{q_\text{min}}{q(x)}}\right]\equiv\rho_- <1/2,\,\label{rho-}\\
                   &=& \frac{1}{2}\left[1+\sqrt{1-\frac{q_\text{min}}{q(x)}}\right]\equiv\rho_+ >1/2.\label{rho+}
\end{eqnarray}
Unsurprisingly, all of $\rho_\text{LD}(x),\rho_\text{HD}(x),\rho_\text{MC}(x)$ depend on the specific choices of the hopping rate function $q(x)$.  Notice that $\rho_\text{LD}(x)$ and $\rho_\text{HD}(x)$, being explicitly dependent on $q(x)$ as given in (\ref{rho-ld}) and (\ref{rho-hd}) above, can be {\em discontinuous}, only when $q(x)$ itself is discontinuous, else they are continuous in $x$.  In contrast, no such statement can be made about $\rho_\text{MC}(x)$, which is generally given by {\em two} distinct functions, as given in (\ref{rho-}) and (\ref{rho+}). Furthermore, at the locations $x_0$ of the minimum of $q(x)$, one has $q(x_0)=q_\text{min}$, and at those points, $\rho_+(x_0)=\rho_-(x_0)=1/2$. This means $\rho_\text{MC}(x_0)=1/2$. These results are generic, independent of any specific form of $q(x)$. We see below that our explicit enumeration of the stationary densities agree with these general conclusions. 

Below we calculate the stationary densities for a few specific forms of $q(x)$. { The choices for $q(x)$ made here either have more than one minimum or have one discontinuity.  These choices complement and extend the studies in Ref.~\cite{atri-jstat}. }  From theoretical perspectives, our various choices for $q(x)$ further demonstrate the problems and (apparent) ambiguities in applying MFT na\"ively in the MC phases, and then show how to get around these difficulties by using physically intuitive arguments.  While these are simple, idealised choices, nonetheless, these can be useful in studies on and inspired by, e.g., rare codon clusters in cells~\cite{cluster}. In addition these choices allow us to connect the analyticity properties of a chosen $q(x)$ with the resulting stationary densities of DDW envelops. { We expect studying these simple choices for $q(x)$ should help to understand the effects for more complex forms for $q(x)$, as may be found in e.g., rare codon
clusters in cells~\cite{cluster}.}


\subsection{Case I}

\begin{figure}[htb]
 \includegraphics[width=8.5cm]{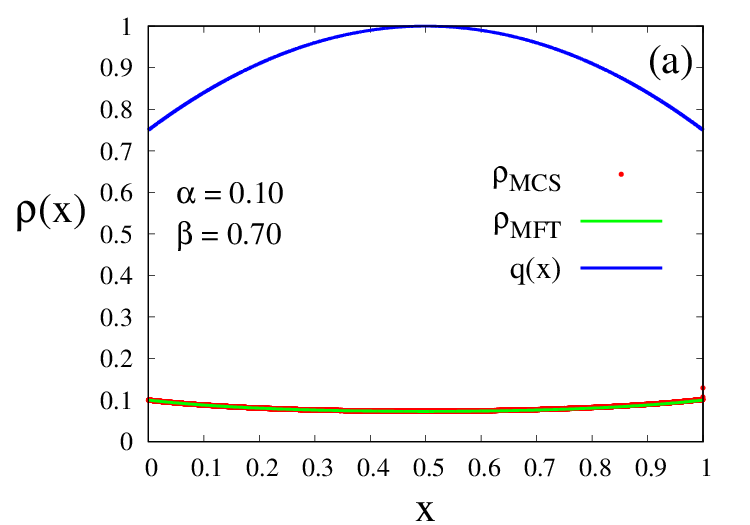}\\
 \includegraphics[width=8.5cm]{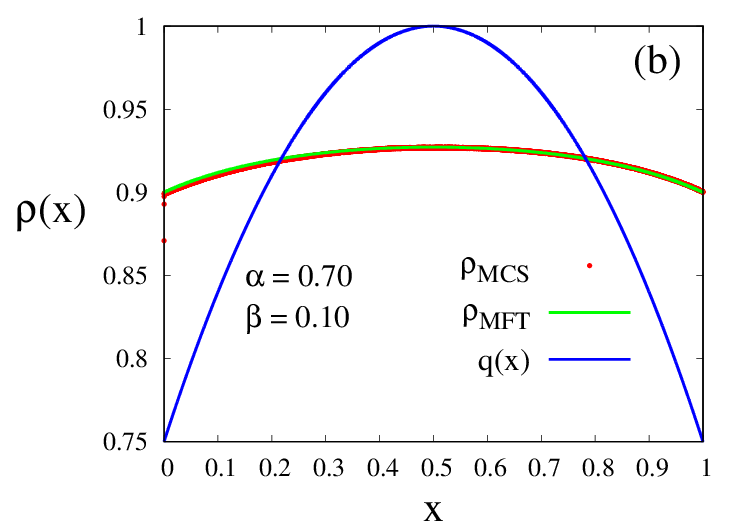}
 \caption{Plots of the stationary densities $\rho(x)$ as a function of $x$ for $q(x)$ given by (\ref{case1}). (a) Density profile $\rho(x)=\rho_\text{LD}(x)$ in the LD phase and (b) density profile $\rho(x)=\rho_\text{HD}(x)$ in the HD phase. The blue curve denotes the hopping rate function $q(x)$. The MCS and MFT results on the density profiles are shown in points (red) and line (green). Good agreements between MFT and MCS results are found (see text).}
 \label{case1-ldhd}
\end{figure}

In Case I, we set
\begin{equation}
 q(x)=1-\left(x-\frac{1}{2}\right)^2.\label{case1}
\end{equation}
Thus, $q(x)$ has a minimum of 3/4 at $x=0$ and $x=1$, and has a maximum of 1 at $x=1/2$. Hence $q(x)$ has {\em two} minima at the two end points of the TASEP lane and a single maximum in the bulk. The stationary densities $\rho_\text{LD}(x)$ and $\rho_\text{HD}(x)$ can be directly obtained from (\ref{rho-ld}) and (\ref{rho-hd}) with $q(x)$ as in (\ref{case1}) and $J=J_\text{LD}$ and $J=J_\text{HD}$, respectively. Our MFT results and the corresponding MCS results for $\rho_\text{LD}(x)$ and $\rho_\text{HD}(x)$ are shown in Fig.~\ref{case1-ldhd}(a) and Fig.~\ref{case1-ldhd}(b) respectively. We find good agreements between MFT and MCS results. 


We now study the MC phase stationary density $\rho_\text{MC}(x)$ that requires careful consideration. In the MC phase, $J=q_\text{min}/4$, which can be used in (\ref{rho-}) or (\ref{rho+}) to construct $\rho_\text{MC}(x)$. Thus the task now is to find out which part of the stationary MC phase density corresponds to $\rho_+(x)$ and which part $\rho_-(x)$.

\begin{figure}[htb]
 \includegraphics[width=8.5cm]{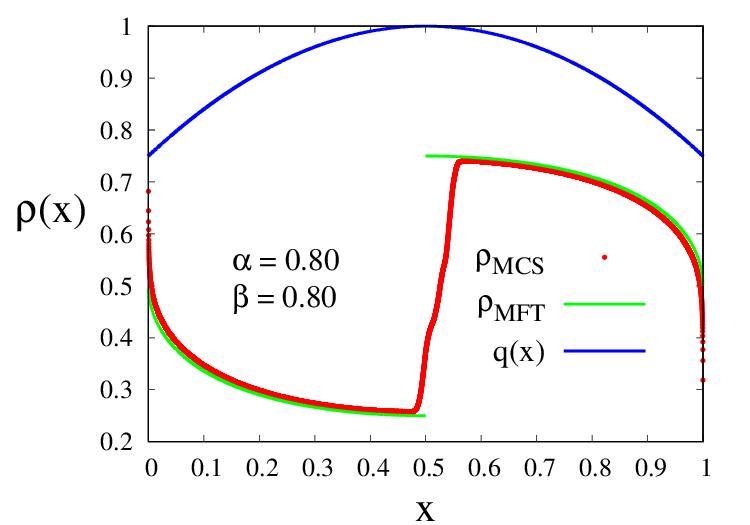}
 \caption{Plot of the stationary density $\rho(x)=\rho_\text{MC}(x)$ in the MC phase for $q(x)$ given by (\ref{case1}). The blue curve denotes the hopping rate function $q(x)$. The MCS and MFT results on the density profiles are shown in points (red) and line (green). Good agreements between MFT and MCS results are found (see text).} \label{case1-mc}
\end{figure}

{ To proceed, we imagine the TASEP lane to be composed of two TASEP lanes $T_\text{L}$ and $T_\text{R}$ connected serially at $x=x_0=1/2$, the location of the maximum of $q(x)$. In this representation, the hopping rate at the exit end of $T_\text{L}$, which is at $x=1/2$, is unity, which is the hopping rate at the entry end of $T_\text{R}$. At $x=0,\,1$, which are the locations of the minima of $q(x)$, both $\rho_+(x),\rho_-(x)= 1/2$. Since $\rho_\text{MC}(x)$ is either $\rho_-(x)$ or $\rho_+(x)$, this means $\rho_\text{MC}(x)=1/2$ {\em definitely} at both $x=0,1$  within our MFT, while $\rho_\text{MC}(x\rightarrow 1/2_\pm)$ is {\em definitely not} 1/2, since neither of $\rho_+(x),\,\rho_-(x)$ is 1/2 there. Let us first focus on $T_\text{L}$. Our MFT gives that it has a stationary density of 1/2 at $x=0$, its entry end, and a different value at $x=1/2$, its exit end. Compare it now with an open, uniform TASEP of size 1/2 between $x=0$ and $x=1/2$, having unit bulk hopping rate with parameter values corresponding its LD-MC phase boundary, i.e., entry rate $ \alpha=1/2$ and exit rate $\beta>1/2$. The corresponding density at $x=0$, the entry end, is 1/2, and at  $x=1/2$, the exit end, is {\em different} from 1/2 (less than 1/2 { in our MCS studies}). { Now arguing to imagine the latter to be a limiting case of $q(x)\rightarrow q$, a constant and} comparing the boundary values and the bulk stationary current $J=q/4=J_\text{max}$, 
we thus see that the density profile in $T_\text{L}$ actually resembles a stationary density found for parameters on the LD-MC phase boundary, in which $J=J_\text{max}=q_\text{min}/{4}$, and has a density 1/2 at the entry end. This argument suggests that between $x=0$ and $x=1/2$, $\rho_\text{MC}(x)=\rho_-(x)$ in the bulk, since $\rho_\text{MC}(x)$ has a density of 1/2 at its entry end $x=0$ and has a value different from 1/2 at $x=1/2$.  
We next consider $T_\text{R}$ and use logic similar to our analysis for $T_\text{L}$ above. Within our MFT, the stationary density on $T_\text{R}$ {\em definitely} has a value 1/2 at $x=1$, its exit end, since $\rho_+(x=1)=\rho_-(x=1)=1/2$, but has a value {\em different from} 1/2 at $x=1/2$, its entry end. Once again compare it with an open, uniform TASEP between $x=1/2$ (entry end) and $x=1$ (exit end) and having unit bulk hopping rate. For parameters corresponding to its HD-MC phase boundary, i.e., exit rate $\beta=1/2$ and entry rate $\alpha>1/2$. The corresponding density at $x=1$, its exit end, is 1/2, and at $x=1/2$ its density is {\em different} from  1/2 (actually more than 1/2 { in our MCS studies}). { In analogy with our arguments for $T_\text{L}$, we thus see that} the steady state density on $T_\text{R}$ resembles the stationary density found for parameters on the HD-MC phase boundary, in which $J=J_\text{max}=q_\text{min}/4$, and has a density 1/2 at its exit end. This argument suggests that between $x=1/2$ and $x=1$, $\rho_\text{MC}(x)=\rho_+(x)$ in the bulk.} We thus arrive at the following MFT solution for $\rho_\text{MC}(x)$:
\begin{eqnarray}
 \rho_\text{MC}(x)&=& \rho_-(x),\,\,\,0\leq x\leq 1/2,\nonumber \\
                  &=& \rho_+(x),\,\,\,1/2\leq x\leq 1.\label{rho-mc-case1}             
\end{eqnarray}
We have plotted (\ref{rho-mc-case1}) together with the corresponding MCS results in Fig.~\ref{case1-mc}. We find excellent agreement of our MFT result on $\rho_\text{MC}(x)$ with the corresponding MCS result. Notice that in this case the spatial average of the MC phase density
\begin{equation}
 \overline \rho_\text{MC}\equiv \int_0^1 \rho_\text{MC}(x)=\frac{1}{2},
\end{equation}
same as the bulk density for the MC phase in a uniform open TASEP. There is a however a fundamental distinction between the two. In a uniform open TASEP, $\rho_\text{MC}(x)=1/2$ {\em everywhere} in the bulk of the TASEP lane. In contrast, in Case I here, $\rho_\text{MC}(x)=1/2$ {\em only} for $x\rightarrow 0,1$. Everywhere else in the bulk, it is {\em different} from 1/2.

\subsection{Case II}

\begin{figure}[htb]
  \includegraphics[width=8.5cm]{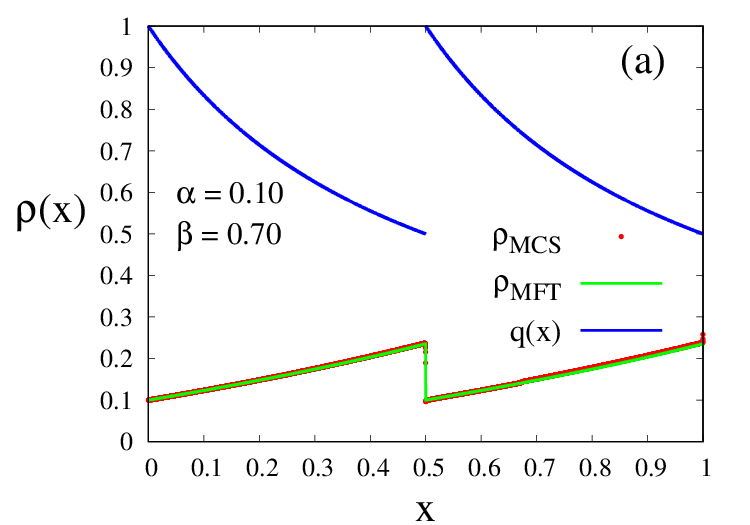}\\
  \includegraphics[width=8.5cm]{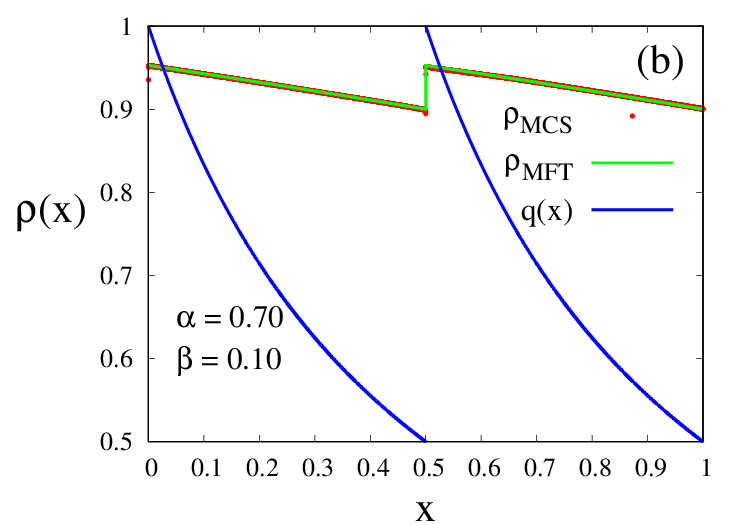}
  \caption{Plots of the stationary densities $\rho(x)$ as a function of $x$ for $q(x)$ given by (\ref{case2}). (a) Density profile $\rho(x)=\rho_\text{LD}(x)$ in the LD phase and (b) density profile $\rho(x)=\rho_\text{HD}(x)$ in the HD phase. The blue curves denote the hopping rate function $q(x)$. The MCS and MFT results on the density profiles are shown in points (red) and line (green). Good agreements between MFT and MCS results are found (see text).}
  \label{case2-ldhd}
 \end{figure}

In Case II, we consider 
\begin{eqnarray}
 q(x) &=& \frac{1}{1+2x},\,0\leq x\leq 1/2,\nonumber \\
      &=& \frac{1}{2x},\,1/2\leq x\leq 1.\label{case2}
\end{eqnarray}
 Thus $q(x)$ is discontinuous at $x=1/2$. It has a maximum value of unity at $x=0$ and $x\rightarrow 1/2_+$, and has minimum value of 1/2 at $x\rightarrow 1/2_-$ and $x=1$. Therefore it has two maxima and two minima. Stationary densities $\rho_\text{LD}(x)$ and $\rho_\text{HD}(x)$ can be straightforwardly obtained from (\ref{rho-ld}) and (\ref{rho-hd}) above together with $q(x)$ given in (\ref{case2}). Unsurprisingly, both $\rho_\text{LD}(x)<1/2$ and $\rho_\text{HD}(x)>1/2$ have discontinuities at $x=1/2$, reflecting the discontinuity of $q(x)$ at that point. Our MFT results for $\rho_\text{LD}(x)$ and $\rho_\text{HD}(x)$ superposed with the corresponding MCS results are shown in Fig.~\ref{case2-ldhd}(a) and Fig.~\ref{case2-ldhd}(b) respectively. We find excellent agreements between the two.


\begin{figure}[htb]
  \includegraphics[width=8.5cm]{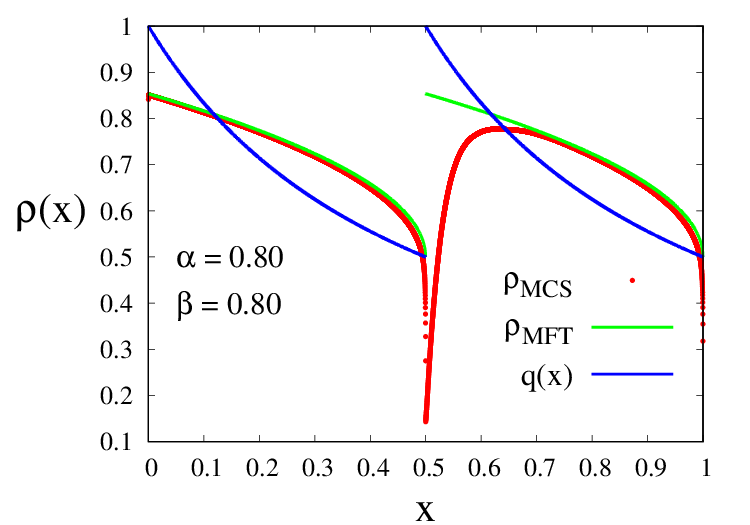}
  \caption{Plot of the stationary density $\rho(x)=\rho_\text{MC}(x)$ in the MC phase for $q(x)$ given by (\ref{case2}). The blue curve denotes the hopping rate function $q(x)$. The MCS and MFT results on the density profiles are shown in points (red) and line (green). Good agreements between MFT and MCS results are found (see text).}
  \label{case2-mc}
 \end{figure}

 We now focus on $\rho_\text{MC}(x)$, the stationary MC phase density. As in the previous case, both $\rho_+(x)$ and $\rho_-(x)$ with $q(x)$ in (\ref{case2}) have a steady current $J=J_\text{max}=q_\text{min}/4$. We must decide whether $\rho_+(x)$ or $\rho_-(x)$ or both should be used to construct $\rho_\text{MC}(x)$ in MFT. We proceed as in the previous case. We imagine the TASEP lane to be composed of two segments $T_\text{L}$ ($0\leq x\leq 1/2$) and $T_\text{R}$ ($1/2\leq x\leq 1$) connected serially at $x=1/2$. In the present case, $q(x_0)=q_\text{min}$ for $x_0= 1/2_-,1$, which gives  $\rho_+(x\rightarrow 1/2_-)=1/2=\rho_-(x\rightarrow 1/2_-)$ and $\rho_+(x\rightarrow 1_-)=1/2=\rho_-(x\rightarrow 1_-)$.
 We use arguments similar to the previous example to determine the stationary densities in the MC phase.
 { Let us focus on $T_\text{L}$ first, which has $x=0$ as its entry end and $x=1/2$ as its exit end. By using our MFT, on $T_\text{L}$, we must have $\rho_\text{MC}(x\rightarrow 1/2_-)=1/2$ as {\em both} $\rho_+(x)$ and $\rho_-(x)$ approach 1/2 at point, whereas $\rho_\text{MC}(x)$ {\em must be different} from 1/2 at $x=0$, since 
 both $\rho_+(x)$ and $\rho_-(x)$ are {\em different} from 1/2 at that point. Thus, on $T_\text{L}$ we have a stationary density that is 1/2 at $x\rightarrow 1/2_-$, the exit end, but is different from 1/2 at $x=0$, its entry end. Once again consider an open TASEP between $x=0$ and 1/2, having unit hopping rate in the bulk. When it has parameter values corresponding to its HD-MC phase boundary, i.e., with exit rate $ \beta=1/2$ and entry rate $\alpha>1/2$, the corresponding stationary density at the exit end $x=1/2$ is 1/2 and at the entry end $x=0$ is {\em different} from (actually more than) half. { As in the previous example, arguing to imagine the latter to be a limiting case of $q(x)\rightarrow q$, a constant and} comparing these boundary values with $T_\text{L}$ having a bulk stationary current $J=q/4=J_\text{max}$, we thus see that the stationary density on $T_\text{L}$ actually resembles the stationary density profile found for parameters on the HD-MC boundary of an open TASEP. We thus expect $\rho_+(x)$ to be the { MFT solution for the} stationary density on the TASEP lane in the range $0\leq x\leq 1/2$. We now consider $T_\text{R}$. We have the stationary density 1/2 at $x=1$, since {\em both} $\rho_+(x)$ and $\rho_-(x)$ are 1/2 at that point, whereas $\rho_\text{MC}(x)$ {\em must be different} from 1/2 at $x\rightarrow 1/2_+$, since 
 both $\rho_+(x)$ and $\rho_-(x)$ are {\em different} from 1/2 at that point. Once again compare with an open TASEP between $x=1/2$ and 1, having a uniform hopping rate in the bulk. By extending arguments made above for $T_\text{L}$ to $T_\text{R}$ and comparing with the boundary values of an open TASEP between $1/2\leq x\leq 1$ with unit hopping rate in the bulk, we conclude that the steady state density profile on $T_\text{R}$ resembles that for parameters on the HD-MC boundary.} Hence, we expect $\rho_+(x)$ to be the stationary density on the TASEP lane in the range $1/2\leq x\leq 1$. We thus conclude that 
 \begin{equation}
  \rho_\text{MC}(x)=\rho_+(x)
 \end{equation}
in the entire TASEP lane, except for the discontinuity at $x=1/2$. Our MFT agrees well with the corresponding MCS result; see Fig.~\ref{case2-mc}.
 Since $\rho_+(x)>1/2$ {\em everywhere} in the TASEP lane except at $x\rightarrow 1/2_-$ and $x=1$, we must have $\overline \rho_\text{MC}>1/2$ in contrast to a uniform open TASEP.

 \subsection{Case III}

 We now consider
 \begin{eqnarray}
  q(x)&=& \frac{1}{2-2x},\,0\leq x\leq 1/2,\nonumber \\
      &=& \frac{1}{3-2x},\,1/2\leq x\leq 1.\label{case3}
 \end{eqnarray}
This is in fact the mirror image of $q(x)$ in (\ref{case2}) about $x=1/2$. Thus $q(x)$ is discontinuous at $x=1/2$. It has two minima of value 1/2 at $x=0$ and $x\rightarrow 1/2_+$, and two maxima at $x\rightarrow 1/2_-$ and $x=1$. Stationary densities $\rho_\text{LD}(x)$ and $\rho_\text{HD}(x)$ can be straightforwardly obtained from (\ref{rho-ld}) and (\ref{rho-hd}) above together with $q(x)$ given in (\ref{case3}). { Due to the particle-hole symmetry, the stationary densities $\rho_\text{LD}(x)$ and $\rho_\text{HD}(x)$ can be read off the stationary high and low density solutions for (\ref{case2}) together with $\alpha\leftrightarrow \beta$. Here, we apriori calculate $\rho_\text{LD}(x)$ and $\rho_\text{HD}(x)$ in our MCS studies for a set of $\alpha,\beta$ not connected to those used in (\ref{case2}) by the particle-hole symmetry (and hence the stationary density being not directly calculable by using the particle-hole symmetry). For $\rho_\text{MC}(x)$ here, the choice of $\alpha,\beta$ are same as those for $q(x)$ is (\ref{case2}) above. Nevertheless, we calculate $\rho_\text{MC}(x)$ here apriori, as that allows us to address the question of how to apply MFT in the MC phase, complementing our analysis for $q(x)$ in (\ref{case2}). We shall however see that our results for $\rho_\text{MC}(x)$ are clearly in agreement with the particle-hole symmetry, which gives further credence to our physical argument for applying MFT in the MC phase. Detailed results on the stationary densities in each of the phases are given below.}

\begin{figure}[htb]
   \includegraphics[width=8.5cm]{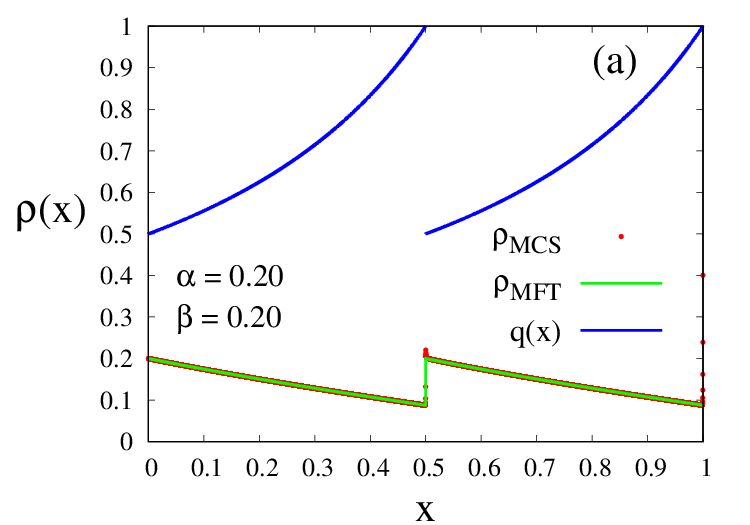}\\
   \includegraphics[width=8.5cm]{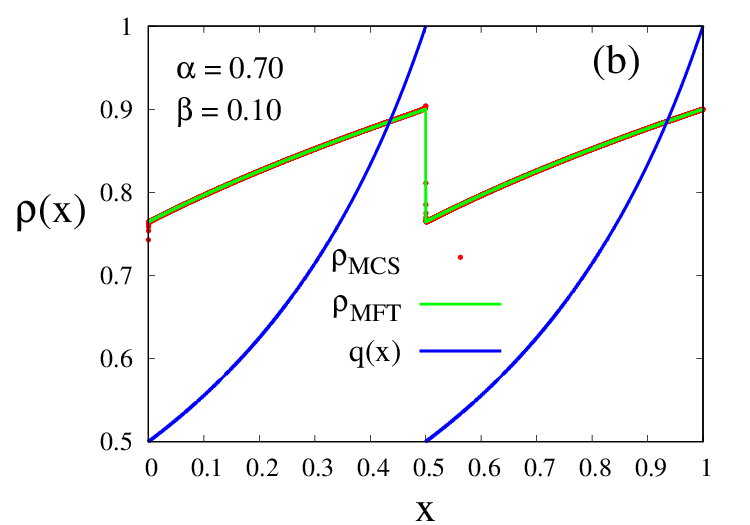}
   \caption{Plots of the stationary densities $\rho(x)$ as a function of $x$ for $q(x)$ given by (\ref{case3}). (a) Density profile $\rho(x)=\rho_\text{LD}(x)$ in the LD phase and (b) density profile $\rho(x)=\rho_\text{HD}(x)$ in the HD phase. The blue curve denotes the hopping rate function $q(x)$. The MCS and MFT results on the density profiles are shown in points (red) and line (green). Good agreements between MFT and MCS results are found (see text).}
   \label{case3-ldhd}
  \end{figure}

Unsurprisingly, both $\rho_\text{LD}(x)<1/2$ and $\rho_\text{HD}(x)>1/2$ have discontinuities at $x=1/2$, reflecting the discontinuity of $q(x)$ at that point. Our MFT results for $\rho_\text{LD}(x)$ and $\rho_\text{HD}(x)$ superposed with the corresponding MCS results are shown in Fig.~\ref{case3-ldhd}(a) and Fig.~\ref{case3-ldhd}(b) respectively. We find excellent agreements between the two.
 
  
\begin{figure}[htb]
  \includegraphics[width=8.5cm]{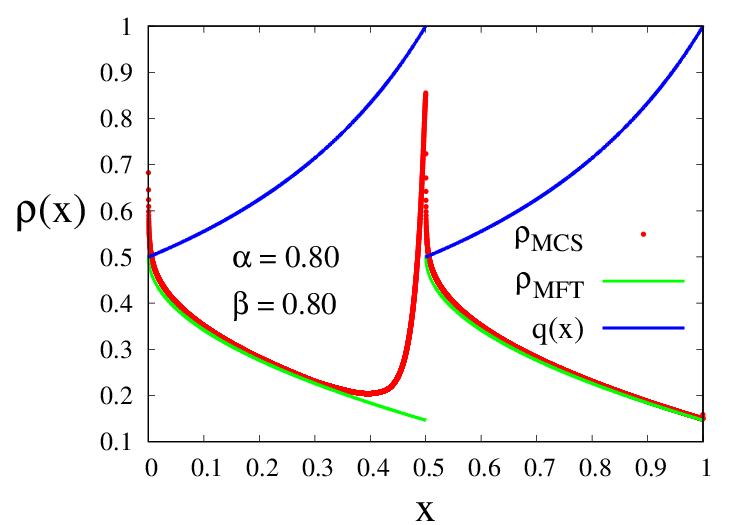}
  \caption{Plot of the stationary density $\rho(x)=\rho_\text{MC}(x)$ in the MC phase for $q(x)$ given by (\ref{case3}). The blue curve denotes the hopping rate function $q(x)$. The MCS and MFT results on the density profiles are shown in points (red) and line (green). Good agreements between MFT and MCS results are found (see text).}
  \label{case3-mc} 
 \end{figure}

 { We now focus on $\rho_\text{MC}(x)$, the stationary MC phase density. As in the previous case, both $\rho_+(x)$ and $\rho_-(x)$ with $q(x)$ in (\ref{case3}) have a steady current $J=J_\text{max}=q_\text{min}/4$. We thus need to decide whether $\rho_+(x)$ or $\rho_-(x)$ or both should be used to construct $\rho_\text{MC}(x)$ in MFT. We proceed as in the previous case. Once again we imagine the TASEP lane to be composed of two segments $T_\text{L}$ ($0\leq x\leq 1/2$) and $T_\text{R}$ ($1/2\leq x\leq 1$) connected serially at $x=1/2$. In the present case, we have $\rho_+(x\rightarrow 0)=1/2=\rho_-(x\rightarrow 0)$ and $\rho_+(x\rightarrow 1/2_+)=1/2=\rho_-(x\rightarrow 1/2_+)$, since $x=0, 1/2_+$ are the locations of the minima of $q(x)$. Let us focus on $T_\text{L}$ first. We use arguments similar to the previous examples. According to our MFT, on $T_\text{L}$, we must have $\rho_\text{MC}(x\rightarrow 0)=1/2$ as {\em both} $\rho_+(x)$ and $\rho_-(x)$ approach 1/2 at point, whereas $\rho_\text{MC}(x)$ {\em must be different} from 1/2 at $x=1/2_-$, since 
 both $\rho_+(x)$ and $\rho_-(x)$ are {\em different} from 1/2 at that point. Thus, on $T_\text{L}$ we have a stationary density that is 1/2 at $x\rightarrow 0$, the entry end, but is different from 1/2 at $x=1/2$, its exit end. Once again consider an open, uniform TASEP between $x=0,1/2$ having unit hopping rate in the bulk. With its entry rate 
 $\alpha=1/2$ and exit rate $\beta>1/2$, it has a density of 1/2 at the entry end $x=0$ and a density {\em different} from 1/2 (actually less than 1/2 { in our MCS studies}) at the exit end $x=1/2$. { Again arguing to imagine the latter to be a limiting case of $q(x)\rightarrow q$, a constant with a stationary current $J=q/4=J_\text{max}$ and} comparing boundary values, we thus see that the stationary density profile resembles the stationary density profile of an open TASEP for parameters on the LD-MC phase boundary. We thus expect $\rho_-(x)$ to be the stationary density on the TASEP lane in the range $0\leq x\leq 1/2$. We now consider $T_\text{R}$. We have a stationary density 1/2 at $x=1/2_+$, since {\em both} $\rho_+(x)$ and $\rho_-(x)$ approach 1/2 at point, whereas $\rho_\text{MC}(x)$ {\em must be different} from 1/2 at $x\rightarrow 1$, since 
 both $\rho_+(x)$ and $\rho_-(x)$ are {\em different} from 1/2 at that point. In analogy with our argument for $T_\text{L}$ above, we argue that  the steady state density profile on $T_\text{R}$ resembles that for parameters on the LD-MC phase boundary. Hence, we expect $\rho_-(x)$ to be the stationary density on the TASEP lane in the range $1/2\leq x\leq 1$.} We thus conclude that 
 \begin{equation}
  \rho_\text{MC}(x)=\rho_-(x)
 \end{equation}
in the entire TASEP lane, except for a discontinuity at $x=1/2$. Our MFT agrees well with the corresponding MCS result; see Fig.~\ref{case3-mc}.
 
Since $\rho_-(x)<1/2$ {\em everywhere} in the TASEP lane except at $x\rightarrow 0$ and $x\rightarrow 1/2_+$, we must have $\overline \rho_\text{MC}<1/2$ in contrast to a uniform open TASEP. { Unsurprisingly, our results for $\rho_\text{MC}(x)$ here are in agreement with what one would obtain by applying the particle-hole symmetry on the corresponding results for $\rho_\text{MC}(x)$ with $q(x)$ in (\ref{case2}), which lends further credence to our calculation for $\rho_\text{MC}(x)$.}
 
 


{ In this Section, we have studied the stationary densities for $q(x)$ having two minima or a discontinuity, or both. This extends the studies reported in Ref.~\cite{atri-jstat}, and suffices to clarify how MFT can still be employed, together with physically intuitive arguments, to correctly obtain   the stationary densities in the MC phases, issues which remained unsettled till date; see, e.g., Ref.~\cite{erdmann} in this context. Stationary densities for more complex choices for $q(x)$, e.g., with more discontinuities and more global minima, can be calculated by extending the calculational framework and intuitive arguments used above.}

\section{Phase transitions and universal phase diagrams }\label{phase}
 
 {  As announced in the beginning, a major goal of the present work is to study the DDWs in open TASEPs, when the hopping rate is nonuniform.  In open uniform TASEPs, DDWs are found only at the LD-HD phase transitions, which are discontinuous in nature, and the DDW envelops are the long time averaged density profiles at the phase coexistence. With spatially nonuniform $q(x)$, even the LD-MC and HD-MC transitions {\em can be} discontinuous in nature. Therefore,  in order to study the DDWs, it is imperative to know which transitions are discontinuous for a given $q(x)$; see Ref.~\cite{atri-jstat} for a related study on this. We revisit this briefly here, calculate the phase diagrams within the MFT, which suffices for our purposes  to identify the discontinuous transitions in straightforward ways. }
 
 Below we calculate the mean-field phase diagrams in the $\alpha-\beta$ plane for a few representative examples of $q(x)$. The different phase boundaries can be obtained by equating the corresponding stationary currents. Using the stationary currents in the LD, HD and MC phases, respectively, as defined in Eqs.~(\ref{j-ld})-(\ref{j-mc}), we obtain the general conditions for obtaining the phase boundaries as follows. The LD-MC phase boundary is given by
 \begin{equation}
  q(0)\alpha(1-\alpha)=\frac{q_\text{min}}{4}.\label{ld-mc-phase}
 \end{equation}
 The  HD-MC phase boundary is given by
 \begin{equation}
  q(1)\beta(1-\beta) = \frac{q_\text{min}}{4}.\label{hd-mc-phase}
 \end{equation}
 If the LD-MC or HD-MC transitions are discontinuous, then DDWs are expected at these transitions with phase boundaries given, respectively, by (\ref{ld-mc-phase}) or (\ref{hd-mc-phase}). 
Lastly, the LD-HD phase boundary is given by
\begin{equation}
 q(0)\alpha(1-\alpha)=q(1)\beta(1-\beta),\label{ld-hd-phase}
\end{equation}
which unsurprisingly is the line in the $\alpha-\beta$ plane for which a DDW is observed.

{ To characterise the phase transitions, we use the order parameters defined in Ref.~\cite{atri-jstat}, which we revisit below. We first define the mean densities in each of the three phases. The mean LD, HD and MC phase densities are 
\begin{eqnarray}
 &&\overline \rho_\text{LD}(x)= \int_0^1 dx \rho_\text{LD}(x),\\
 &&\overline \rho_\text{HD}(x)= \int_0^1 dx \rho_\text{HD}(x),\\
 &&\overline \rho_\text{MC}(x)= \int_0^1 dx \rho_\text{MC}(x).
 \end{eqnarray}
 We then define order parameters~\cite{atri-jstat}
 \begin{eqnarray}
 &&\Delta\rho_\text{LD-HD}= \overline \rho_\text{HD}-\overline\rho_\text{LD},\\
 &&\Delta\rho_\text{LD-MC}= \overline \rho_\text{MC}-\overline\rho_\text{LD},\\
 && \Delta\rho_\text{HD-MC}= \overline \rho_\text{MC}-\overline\rho_\text{HD}.
 \end{eqnarray}
Clearly, by using (\ref{rho-ld}) and (\ref{rho-hd}),  $\Delta\rho_\text{LD-HD}>0$ at the LD-HD transition {\em always}, which implies a discontinuous LD-HD  transition {\em always}. On the other hand, we will see that $ \Delta\rho_\text{LD-MC}$ and $\Delta\rho_\text{HD-MC}$ can be zero or nonzero, corresponding to continuous or discontinuous transitions. 

}

We now explicitly calculate the phase boundaries for some choices of $q(x)$. \\

\begin{figure}[htb]
 \includegraphics[width=8.7cm]{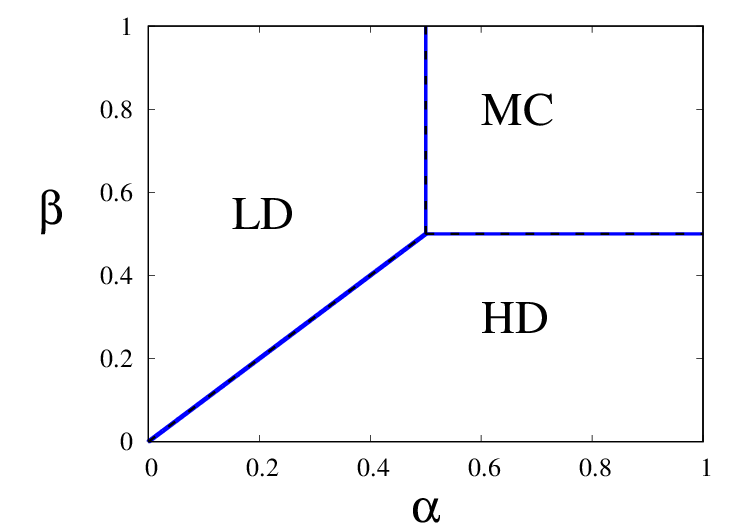}
  \caption{Phase diagram in the $\alpha-\beta$ plane with $q(x)$ as given by (\ref{case1}) above. It is identical to the phase diagram for a uniform open TASEP. However, all the transitions are discontinuous in nature; see text.} \label{phase-diag1}
 \end{figure}

(i) We begin with Case I above. With $q(x)$ as given in (\ref{case1}), we note that 
\begin{equation}
 q(0)=q_\text{min}=q(1)=\frac{3}{4}.\label{case1-qphase}
\end{equation}
This gives by using (\ref{ld-mc-phase})
\begin{equation}
 \alpha(1-\alpha)=\frac{1}{4} \implies \alpha=\frac{1}{2} \label{ld-mc-1}
\end{equation}
as the LD-MC phase boundary. Similarly, by using (\ref{hd-mc-phase})
\begin{equation}
 \beta(1-\beta)=\frac{1}{4}\implies \beta = \frac{1}{2} \label{hd-mc-1}
\end{equation}
as the HD-MC phase boundary. Finally, we get
\begin{equation}
 \alpha(1-\alpha)=\beta(1-\beta)\leq \frac{1}{4} \implies \alpha = \beta\leq \frac{1}{2} \label{ld-hd-1}
\end{equation}
is the phase boundary between the LD and HD phases.  We thus find that the phase boundaries (\ref{ld-mc-1}), (\ref{hd-mc-1}) and (\ref{ld-hd-1}) are {\em identical} to their counterparts for a uniform open TASEP, with the phase boundaries meeting at (1/2,1/2); see Fig.~\ref{phase-diag1}. Thus the phase diagram in the $\alpha-\beta$ plane is identical with that for a uniform open TASEP. 
Thus calculation of the phase diagram {\em cannot} distinguish between a uniform open TASEP and an open TASEP with a hopping rate function given in (\ref{case1}). Are these two cases statistically identical then? We argue that this is not so. In fact, our Case I is fundamentally different from an open uniform TASEP. This is revealed in the nature of the respective phase transitions in the two models. In case of a uniform open TASEP, the LD-MC and HD-MC transitions are continuous transition~\cite{derrida-rev,blythe}, since the LD phase bulk density $\alpha$  smoothly matches with the MC phase bulk density of 1/2 at $\alpha=1/2$. Similar considerations for the HD to MC transition show that it is also a continuous transition. In contrast, in Case I of our study, $\rho_\text{LD}(x)<1/2$ {\em even} at the transition point to the MC phase with $J_\text{LD}=J_\text{MC}$, except at isolated points where $q(x)=q_\text{min}$. This means the spatial average of the stationary LD phase density 
\begin{equation}
\overline \rho_\text{LD}(x)= \int_0^1 dx \rho_\text{LD}(x) <\overline \rho_\text{MC}= \int_0^1dx \rho_\text{MC}(x)=\frac{1}{2},\label{ld-mean}
\end{equation}
the mean density in the MC phase, even at the the transition point,  giving a {\em discontinuous transition}, in contrast to a uniform open TASEP; see also Refs.~\cite{atri-jstat,erdmann}. 
 The jump in the mean densities across the LD-MC transition is given by
\begin{equation}
 \Delta\rho_\text{LD-MC}= \overline \rho_\text{MC}-\overline\rho_\text{LD}= \frac{1}{2}\int_0^1 dx\,\sqrt{1-\frac{q_\text{min}}{q(x)}}.\label{ld-mc-jump}
\end{equation}

Similar arguments as above give that the HD-MC transition is also a discontinuous transition: the spatial average of the HD phase density 
\begin{equation}
\overline \rho_\text{HD}= \int_0^1 dx \rho_\text{HD}(x)>\overline\rho_\text{MC}=1/2,\label{hd-mean}
\end{equation}
since $\rho_\text{HD}(x)>1/2$ {\em even} at the transition point to the MC phase with $J_\text{HD}=J_\text{MC}$, except at isolated points where $q(x)=q_\text{min}$.  The jump in the mean densities across the LD-MC transition is given by
\begin{equation}
 \Delta\rho_\text{HD-MC}= \overline \rho_\text{HD}-\overline\rho_\text{MC}= \frac{1}{2}\int_0^1 dx\,\sqrt{1-\frac{q_\text{min}}{q(x)}},\label{hd-mc-jump}
\end{equation}
which in this particular case is same as (\ref{ld-mc-jump}).

Lastly, the LD-HD transition remains a discontinuous transition, as in a uniform open TASEP.  The jump in the mean densities across the LD-HD transition is given by
\begin{equation}
 \Delta\rho_\text{LD-HD}= \overline \rho_\text{HD}-\overline\rho_\text{LD}= \int_0^1 dx\,\sqrt{1-\frac{q_\text{min}}{q(x)}},\label{ld-hd-jump}
\end{equation}
which is twice that of (\ref{ld-mc-jump}). The specific relations between the density jumps across the phase boundaries as given in (\ref{ld-mc-jump})-(\ref{ld-hd-jump}) apply only in this case due to the result $\overline \rho_\text{MC}=1/2$. 

Thus all the three transitions are discontinuous transitions here. This means the meeting point (1/2,1/2) of the three phase boundaries in the $\alpha-\beta$ plane is {\em not} a multicritical point, since none of the three transitions here involves any critical point, but is like a triple point in an equilibrium solid fluid system, where solid-liquid, solid-gas and liquid gas transition lines - all first order - meet, i.e., the the point where all the three phases of solid, liquid and gas coexist~\cite{zemansky}!  In the present case, the LD, HD and MC phases coexist at (1/2,1/2).  This has another implication. It means the domain wall height $\Delta(x)$, which depends upon $x$, in general {\em does not} vanish with $\alpha,\beta$ corresponding to  the common meeting point of the phase boundaries for any $x$, except at isolated points where $q(x)=q_\text{min}$ (in this case at $x=0,\,1$). 
In contrast, the domain wall height in a uniform open TASEP necessarily vanishes at all points in the TASEP lane with $\alpha,\beta$ fixed at the multicritical point (1/2,1/2).

 (ii) We now construct the phase diagram for $q(x)$ given in (\ref{case2}). In this case, $q_\text{min}=1/2, q(0)=1, q(1)=1/2$. This gives for the LD-MC phase boundary by using (\ref{ld-mc-phase})
\begin{equation}
 \alpha(1-\alpha)=\frac{1}{8}\implies \alpha=\frac{1}{2}\left[1  - \frac{1}{\sqrt 2}\right]<1/2\label{ld-mc-2}
\end{equation}
as the phase boundary between the LD and MC phases, which is {\em different} from a uniform open TASEP. In contrast, by using (\ref{hd-mc-phase}), we find
\begin{equation}
 \beta(1-\beta)=\frac{1}{4} \implies \beta=\frac{1}{2} \label{hd-mc-2}
\end{equation}
as the phase boundary between the HD and MC phases. This is unchanged from its uniform open TASEP result. The LD-HD phase boundary is given by
\begin{equation}
 \alpha(1-\alpha)=\frac{1}{2}\beta(1-\beta)\implies \beta=\frac{1}{2}\left[1-\sqrt{1-8\alpha(1-\alpha)}\right],\label{ld-hd-2}
\end{equation}
which is a curved line, identical to the coexistence line on which a DDW can be found and clearly different from its counterpart in a uniform open TASEP. Phase boundaries (\ref{ld-mc-2}), (\ref{hd-mc-2}) and (\ref{ld-hd-2}) meet at a common point $[(1-1/\sqrt 2)/2, 1/2]$, which is different from its counterpart in a uniform open TASEP.

\begin{figure}[htb]
 \includegraphics[width=8.7cm]{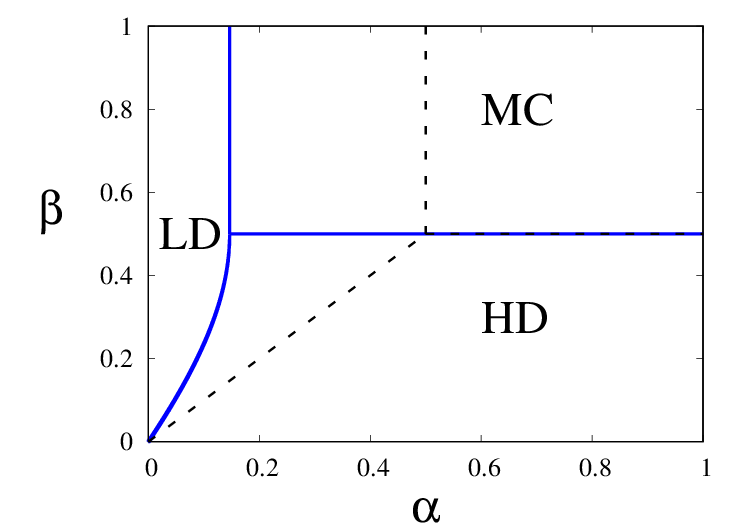}
  \caption{Phase diagram in the $\alpha-\beta$ plane with $q(x)$ as given by (\ref{case2}) above. The phase diagram is different from but has the same topology as that in a uniform open TASEP. The broken lines give the LD-HD and LD-MC phase boundaries in an open, uniform TASEP as reference phase boundaries. The HD-MC phase boundary in the present case overlaps with that in an open, uniform TASEP. The LD-HD and LD-MC transitions are discontinuous, but the HD-MC translation is a continuous transition. See text. }
  \label{phase-diag2}
 \end{figure}

See Fig.~\ref{phase-diag2} for the phase diagram. While it has the same topology as the phase diagram in a uniform open TASEP, the LD-HD and LD-MC phase boundaries have shifted towards the $\beta$-axis, squeezing the LD phase area in the $\alpha-\beta$ plane. The HD-MC phase boundary remains at $\beta=1/2$, but now extends to $[(1-1/\sqrt 2)/2, 1/2]$ towards the $\beta$-axis. 

The LD to HD phase transition is as usual a discontinuous transition, marked by a finite density difference across the phase boundary. The LD to MC phase transition is also a discontinuous transition here for reasons similar to the discontinuous LD to MC phase transition with $q(x)$ in (\ref{case1}) as discussed above. Recall that in this case, $\rho_\text{MC}(x)$ is entirely given by $\rho_+(x)>1/2$ [see (\ref{rho+})], with $J=J_\text{MC}=\frac{q_\text{min}}{4}$, for all $x$. Furthermore, $\rho_\text{LD}(x)=\rho_-(x)<1/2$ with $J=J_\text{LD}<J_\text{MC}$, for all $x$. This means $\overline \rho_\text{LD}<1/2$ even at the threshold of the transition to the MC phase, i.e., when $J_\text{LD}=J_\text{MC}$, whereas $\overline \rho_\text{MC}>1/2$. The jump in the mean densities across the LD-MC phase boundary is again given by (\ref{ld-mc-jump}).
In contrast, the transition between the HD and MC phases remains a continuous transition, as in a uniform open TASEP, albeit at a mean density different from 1/2 at the transition point. This may be seen as follows. Noting that $\rho_\text{MC}(x)$ is entirely given by $\rho_+(x)$ with $J=q_\text{min}/4$, $\rho_\text{HD}(x)$ smoothly matches with $\rho_\text{MC}(x)$ as $J_\text{HD}\rightarrow J_\text{MC}$. Naturally, at the threshold $\overline \rho_\text{HD}= \overline \rho_\text{MC}$, giving a continuous transition. Thus, the common meeting point $[(1-1/\sqrt 2)/2, 1/2]$ of the three phases in the $\alpha-\beta$ plane is a multicritical point {\em different} from its counterpart in an open uniform TASEP: in the present case, it is a meeting point two discontinuous and one continuous transitions, whereas in an open uniform TASEP, two continuous and one discontinuous transition lines meet at (1/2, 1/2). { Close to the HD-MC phase transition, treating $\beta$ as the tuning parameter and $\beta=1/2$ as its critical value, we can define an order parameter exponent as follows. Close to the transition, $\beta=1/2 -\delta$. For $\delta>0$, HD phase ensures. Then,
\begin{eqnarray}
 \Delta\rho_\text{HD-MC}&=&\frac{1}{2}\int_0^1 dx\sqrt{1-\frac{4q(1)(1/2+\delta)(1/2-\delta)}{q(x)}} \nonumber \\&-& \frac{1}{2}\int_0^1 dx\sqrt{1-\frac{q_\text{min}}{q(x)}}\sim \delta^2\times {\cal O}(1),\label{order-expo}
\end{eqnarray}
giving a mean-field order parameter exponent $\tilde\beta= 2$. This is to be contrasted with the open, uniform TASEP case, where
\begin{equation}
 \Delta\rho_\text{HD-MC}=1-\beta -\frac{1}{2}=1-\frac{1}{2}+\delta -\frac{1}{2}=\delta,
\end{equation}
giving $\tilde\beta=1$, different from what we find in the present problem. Therefore, even when there is continuous transition in the present study, it belongs to a universality class different from the open, uniform TASEP case.}

 (iii) { We now construct the phase diagram for $q(x)$ in (\ref{case3}), which is connected with the phase diagram Fig.~\ref{phase-diag2} by particle-hole symmetry, i.e., its phase boundaries can be calculated by interchanging $\alpha\leftrightarrow\beta$ of the phase diagram Fig.~\ref{phase-diag2}. Explicit calculations of the phase diagrams without using the particle-hole symmetry indeed support these predictions, which are presented below.}
 
 With $q(x)$ in (\ref{case3}), $q_\text{min}=1/2, q(0)=1/2, q(1)=1$. This gives for the LD-MC phase boundary by using (\ref{ld-mc-phase})
\begin{equation}
 \alpha(1-\alpha)=\frac{1}{4}\implies \alpha =\frac{1}{2}\label{ld-mc-3}
 \end{equation}
 as the phase boundary between the LD and MC phases. This is unchanged from its uniform open TASEP result. In contrast, by using (\ref{hd-mc-phase}), we find
\begin{equation}
 \beta(1-\beta)=\frac{1}{8} \implies \beta=\frac{1}{2}\left[1-\frac{1}{\sqrt 2}\right] \label{hd-mc-3}
\end{equation}
as the phase boundary between the HD and MC phases. This is a straight line, which is however {\em different} from a uniform open TASEP.  The LD-HD phase boundary is given by
\begin{equation}
 \frac{1}{2}\alpha(1-\alpha)=\beta(1-\beta)\implies \alpha = \frac{1}{2}\left[1-\sqrt{1-8\beta(1-\beta)}\right], \label{ld-hd-3}
\end{equation}
which is a curved line, identical to the coexistence line on which a DDW can be found and clearly different from its counterpart in a uniform open TASEP. Phase boundaries (\ref{ld-mc-3}), (\ref{hd-mc-3}) and (\ref{ld-hd-3}) meet at a common point $[1/2, (1-1/\sqrt 2)/2]$, which is different from its counterpart in a uniform open TASEP.

\begin{figure}[htb]
 \includegraphics[width=8.7cm]{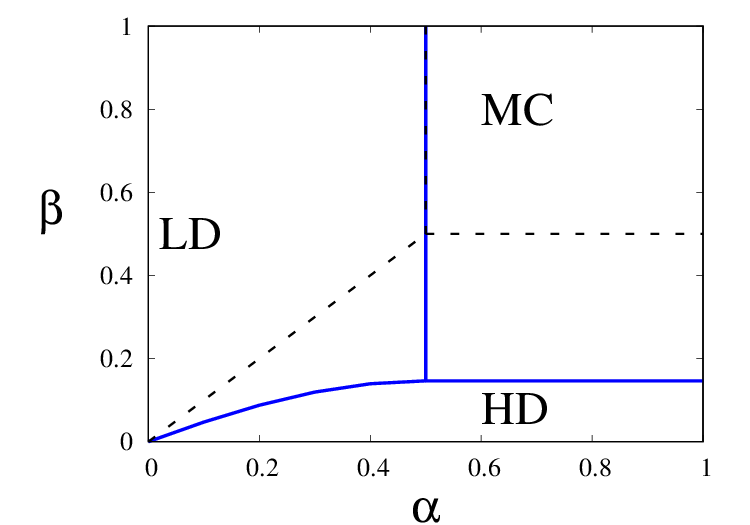}
  \caption{Phase diagram in the $\alpha-\beta$ plane with $q(x)$ as given by (\ref{case3}) above. The phase diagram is different from but has the same topology as that in a uniform open TASEP. The broken lines give the LD-HD and HD-MC phase boundaries in an open, uniform TASEP as reference phase boundaries. The LD-MC phase boundary in the present case overlaps with that in an open, uniform TASEP. The LD-HD and HD-MC transitions are discontinuous, but the LD-MC translation is a continuous transition. See text.}
  \label{phase-diag3}
 \end{figure}

 See Fig.~\ref{phase-diag3} for the phase diagram. While it has the same topology as the phase diagram in a uniform open TASEP, the LD-HD and HD-MC phase boundaries have shifted towards the $\alpha$-axis, squeezing the HD phase area in the $\alpha-\beta$ plane. The LD-MC phase boundary remains at $\alpha=1/2$, but now extends to $[1/2,(1-1/\sqrt 2)/2]$ towards the $\alpha$-axis. 

The LD to HD phase transition is as usual a discontinuous transition, marked by a finite density difference across the phase boundary. 
The HD to MC phase transition is also a discontinuous transition here for reasons similar to the discontinuous HD to MC phase transition with $q(x)$ in (\ref{case1}) as discussed above. Recall that in this case, $\rho_\text{MC}(x)$ is entirely given by $\rho_-(x)<1/2$ [see (\ref{rho+})], with $J=J_\text{MC}=\frac{q_\text{min}}{4}$, for all $x$. Furthermore, $\rho_\text{HD}(x)=\rho_+(x)>1/2$ with $J=J_\text{HD}<J_\text{MC}$, for all $x$. This means $\overline \rho_\text{HD}>1/2$ even at the threshold of the transition to the MC phase, i.e., when $J_\text{LD}=J_\text{MC}$, whereas $\overline \rho_\text{MC}<1/2$. The jump in the mean densities across the HD-MC transition is given by (\ref{hd-mc-jump}) above.

In contrast, the transition between the LD and MC phases remains a continuous transition, same as in a uniform open TASEP (albeit at a mean density different from 1/2 at the transition point). This may be seen as follows. Noting that $\rho_\text{MC}(x)$ is entirely given by $\rho_-(x)$ with $J=q_\text{min}/4$, $\rho_\text{LD}(x)$ smoothly matches with $\rho_\text{MC}(x)$ as $J_\text{LD}\rightarrow J_\text{MC}$. Naturally, at the threshold $\overline \rho_\text{LD}= \overline \rho_\text{MC}$, giving a continuous transition. Notice the mutually complementary picture of the phase diagram and the phase transitions between the last two cases. This is because the inherent complementary nature between the hopping rate function $q(x)$ in these two cases. { Indeed as mentioned above, the phase boundaries (\ref{ld-mc-3}), (\ref{hd-mc-3}) and (\ref{ld-hd-3}) can be obtained from those in the phase diagram (\ref{phase-diag2}) by using the particle-hole symmetry.}

The meeting point $[1/2, (1-1/\sqrt 2)/2]$ is a multicritical point, where two discontinuous and one continuous transition lines meet, different from its counterpart in an open uniform TASEP. { Lastly, the mean-field order parameter exponent $\tilde\beta =2$ across the LD-MC phase continuous transition, which can be shown by setting $\alpha=1/2-\delta$ and expanding $\Delta\rho_\text{LD-MC}$ to the lowest nontrivial order in $\delta$ [see also \eqref{order-expo} above]. Clearly, this is different from its counterpart in open, uniform TASEP, establishing a distinct universality class.}

 (iv) We now consider the phase diagram for 
 \begin{equation}
  q(x)=\frac{1}{1+x}.\label{qx-ddw1}
 \end{equation}
Thus, $q(0)=1,\,q(1)=1/2=q_\text{min}$. Following the logic outlined above, the LD-HD phase boundary is given by
\begin{equation}
 J_\text{LD}=J_\text{HD}\implies \beta=\frac{1}{2}\left[1-\sqrt{1-8\alpha(1-\alpha)}\right].
\end{equation}
The LD-MC phase boundary is given by
\begin{equation}
 J_\text{MC}=J_\text{LD}=\frac{q_\text{min}}{4}\implies \alpha= \frac{1}{2}\left[1-\frac{1}{\sqrt 2}\right].
\end{equation}
Stationary densities $\rho_\text{LD}(x)$ and $\rho_\text{HD}(x)$ are given by (\ref{rho-ld}) and (\ref{rho-hd}). MC phase density $\rho_\text{MC}(x)$ should be a combination of $\rho_\text{LD}(x,J_\text{LD}=q_\text{min}/4)$ and $\rho_\text{HD}(x,J_\text{HD}=q_\text{min}/4)$. Following the logic for $q(x)$ given in (\ref{case1}), we can conclude that in the present case 
\begin{equation}
 \rho_\text{MC}(x)= \rho_\text{HD}(x,J_\text{HD}=q_\text{min}/4) \;\text{ for}\;0\leq x\leq 1.\label{rho-mc-ddw1}
\end{equation}
This also means $\rho_\text{MC}(x)\neq \rho_\text{LD}(x,J_\text{LD}=q_\text{min}/4$. This in turn means at the LD-MC transition, there should be a coexistence of two distinct stationary density solutions $\rho_\text{MC}(x)= \rho_\text{HD}(x,J_\text{HD}=q_\text{min}/4)$ and $\rho_\text{LD}(x,J_\text{LD}=q_\text{min}/4$ with a jump in the order parameter $\Delta\rho_\text{LD-MC}$ at the transition, giving a  DDW at this phase coexistence. In contrast, $\Delta\rho_\text{HD-MC}$  vanishes at the HD-MC transition. The phase diagram for $q(x)$ in (\ref{qx-ddw1}) is identical to the one given in Fig.~\ref{phase-diag2}.

(v) Next we consider 
\begin{equation}
 q(x)=\frac{1}{2-x},\label{qx-ddw2}
\end{equation}
such that $q(0)=1/2=q_\text{min},\,q(1)=1$. Notice that (\ref{qx-ddw2}) is the particle-hole symmetric counterpart of (\ref{qx-ddw1}). In particular, 
\begin{equation}
 \rho_\text{MC}(x)= \rho_\text{LD}(x,J_\text{LD}=q_\text{min}/4) \;\text{ for}\;0\leq x\leq 1.\label{rho-mc-ddw2}
\end{equation}
We can then immediately conclude that the phase diagram is identical to the one in Fig.~\ref{phase-diag3} [which is the particle-hole symmetric counterpart of the phase diagram in Fig.~\ref{phase-diag2}], the LD-MC transition is a continuous transition and the HD-MC transition is a discontinuous one with a DDW at the transition.

(vi) Finally, we consider
\begin{eqnarray}
 q(x)&=& \frac{1}{1+2x},\;\;\;0\leq x\leq 1/2,\nonumber \\
 &=& \frac{1}{3-2x},\;\;\;1/2\leq x\leq 1.\label{qx-ddw3}
\end{eqnarray}
The stationary densities and phase diagrams were originally studied for $q(x)$ in (\ref{qx-ddw3}) by using MFT and MCS in Ref.~\cite{atri-jstat}, which we briefly revisit here.
In this case $q(0)=1=q(1)$, $q(1/2)=1/2=q_\text{min}$. Thus, $q(x)$ in (\ref{qx-ddw3}) has its minimum within the bulk, unlike (\ref{qx-ddw1}) and (\ref{qx-ddw2}), where the minima are at the extreme ends.  It was found that all the three transitions, {\em viz.}, LD-HD, LD-MC and HD-MC are discontinuous. It was also shown that 
\begin{eqnarray}
 \rho_\text{MC}(x) &=& \rho_\text{HD}(x,J_\text{HD}=q_\text{min}/4),\;\;\;0\leq x\leq 1/2\nonumber \\
                   &=& \rho_\text{LD}(x,J_\text{LD}=q_\text{min}/4),\;\;\;1/2\leq x\leq 1, \label{rho-mc-ddw3}
\end{eqnarray}
by combining MCS, and MFT together with physical arguments.
 
\section{Delocalised domain walls}\label{ddw}

{ In Section~\ref{phase} above, we have found that not only the LD-HD phase transition is {\em always} discontinuous, similar to an open TASEP with uniform hopping, even the LD-MC or HD-MC transitions {\em can} also be discontinuous, depending upon the choices for $q(x)$, in contrast to an open, uniform TASEP. A discontinuous transition is necessarily accompanied by coexistence of the two associated phases separated by the discontinuous transition. This coexistence is reflected in the form of a DDW at the LD-HD phase transition in an open, uniform TASEP. Similar reasoning suggests that in the present study a DDW should form not only at the LD-HD transition, but can also form at the LD-MC or HD-MC transition, if these are discontinuous transitions. The surprising fact is that such DDWs at the   LD-MC or HD-MC transition can be full or partial, a novelty not found for the DDW at the LD-HD phase boundary. Lastly and most intriguingly, at the triple point, at which the LD-HD, LD-MC and HD-MC phase boundaries meet, the DDW when ``full'' can actually be interpreted as two separate DDWs connected serially along the TASEP channel. Our numerical simulations confirm these results. Below we calculate the DDW envelops in the different cases.}

We consider a DDW in a TASEP lane with a spatially nonuniform hopping rate function $q(x)$. A DDW in an open TASEP with uniform hopping rate is formed for $\alpha,\beta$ on the coexistence line $\alpha=\beta<1/2$, which is the LD-HD phase boundary  in the $\alpha-\beta$ plane~\cite{kolom}. In the present study, where a DDW can be found at the LD-HD, LD-MC and HD-MC phase transitions, the conditions of a DDW are naturally generalised to
\begin{eqnarray}
 &&J_\text{LD}= q(0)\alpha(1-\alpha)\equiv J_\text{HD}=q(1)\beta(1-\beta)<J_\text{MC}= \frac{q_\text{min}}{4}, \label{coex}\\
 &&J_\text{LD}= q(0)\alpha(1-\alpha)\equiv J_\text{MC}=\frac{q_\text{min}}{4},\label{coex-ldmc}\\
 &&J_\text{HD}= q(0)\alpha(1-\alpha)\equiv J_\text{MC}=\frac{q_\text{min}}{4}.\label{coex-hdmc}
\end{eqnarray}


As in a uniform open TASEP, a DDW in the present model is actually the envelope of a series of localised doman walls (LDWs). Since a DDW is necessarily fluctuating, in order to calculate its envelope, we need to go beyond MFT. Let $x_w$ be the position of an LDW with height $\Delta(x_w)$, which is nothing but the height difference between the two coexisting stationary density solutions. Unlike in an open uniform TASEP, $\Delta$ now depends explicitly on $x_w$, due to the spatially nonuniform hopping rates. Also unlike in an open uniform TASEP, $\Delta$ now be nonzero at the LD-MC or HD-MC transitions.

If we increase the number of particle in the TASEP lane by one, i.e., if one particle enters the TASEP lane through the left (entry) end, the LDW position $x_w$ shifts by an amount $-1/[L\Delta (x_w)]$ (to the left)~\cite{tobias-ef2,tirtha-prr}. Similarly, when one particle leaves the TASEP lane through the right (exit) end, the LDW shifts by an amount $1/[L\Delta(x_w)]$ to the right. Thus there are two basic microscopic processes, which alter the particle number of the TASEP lane and make the LDW shift position { by $\delta x_w$}:\\

(i) One particle entering through the left boundary of the TASEP lane. The corresponding transition rate is given by


$W_L= J_\text{in},\,\delta x_w = - L^{-1}\Delta^{-1}$.
\\

(ii) One particle exiting through the right boundary of the TASEP lane. The corresponding transition rate is given by


$W_R= J_\text{out}, \,\delta x_w = - L^{-1}\Delta^{-1}$.

{ In writing the above forms for $W_L$ and $W_R$, i.e., $W_L$ and $W_R$ are just $J_\text{in}$ and $J_\text{out}$, we are guided by the physically intuitive picture that these transition rates should be proportional to the respective currents (as above) up to a proportionality constant, which has been set to unity by setting the time scale.}

Let $P(x_w,t)$ be the probability of finding a DW at $x_w$ in the TASEP lane at time $t$. The master equation that $P$ satisfies has the general form
\begin{eqnarray}
 \frac{\partial P}{\partial t}&=& \sum_{\delta x_w}P(x_w+\delta x_w,t)W(x_w+\delta x_w\rightarrow x_w) \nonumber \\&-& \sum_{\delta x_w}P(x_w,t)W(x_w\rightarrow x_w+\delta x_w).\label{master-eq}
\end{eqnarray}

By using the different transition rates defined above, we can calculate the average shift or the expectation value of the change $\delta x_w$, given by
\begin{eqnarray}
 \langle \delta x_w\rangle &=& \sum[\text{signed increment}\,\delta x_w\times \text{transition rates}]\nonumber \\ &=& 
 \frac{1}{L\Delta(x_w)}[J_\text{out} -J_\text{in}] =0\nonumber \\
 \label{deloc-cond}
\end{eqnarray}
at the delocalisation transition. This gives
\begin{equation}
J_\text{out}=J_\text{in}\label{deloc-line}
\end{equation}
as the condition for  delocalisation that in simple terms just means ``injection limited current = extraction limited current''. For the LD-HD transition,
\begin{equation}
 J_\text{in}=q(0)\alpha(1-\alpha),\;J_\text{out}=q(1)\beta(1-\beta),\label{deloc-line-ld-hd}
\end{equation}
which together with (\ref{deloc-line}) generalise the well-known $\alpha=\beta$ condition for a DDW at the LD-HD transition in an open uniform TASEP~\cite{blythe}, as mentioned above. In general, with $q(0)\neq q(1)$, the delocalisation line (\ref{deloc-line}) is a curved line for the LD-HD transition. This intriguingly means that even when $q(x)$ is explicitly $x$-dependent, so long as $q(0)=q(1)$, a DDW at the LD-HD transition is observed for $\alpha=\beta$ from (\ref{deloc-line}) above, identical to an open TASEP with a uniform hopping rate. 

Noting that the series of LDWs that make up a DDW necessarily connects coexistence of a lower density segment at the entry end with a higher density  segment at the entry end, we must have for the LD-MC transition
\begin{equation}
 J_\text{in}=q(0)\alpha(1-\alpha),\;J_\text{out} = \frac{q_\text{min}}{4},\label{deloc-line-ld-mc}
\end{equation}
and for the HD-MC transition
\begin{equation}
 J_\text{in} = \frac{q_\text{min}}{4},\;J_\text{out}=q(1)\beta(1-\beta). \label{deloc-line-hd-mc}
\end{equation}
We also note that the delocalisation line (\ref{deloc-line}) for the LD-MC and HD-MC transitions has no analogue in the open, uniform TASEP case, as these transitions in an open, uniform TASEP is {\em always} continuous.

 To proceed further, we note that if $\overline x_w$ is the mean position of the DDW, then $\delta x_w= x_w-\overline x_w$ is the fluctuation around the mean. We now expand  the right hand side of the master equation (\ref{master-eq}) around the mean $\overline x_w$ up to second order in $\delta x_w$. We obtain

\begin{widetext}
 \begin{eqnarray}
  \frac{\partial P}{\partial t}&=& \sum_{\delta x_w}P(x_w+\delta x_w,t)W(x_w+\delta x_w\rightarrow x_w)- \sum_{\delta x_w} P(x_w,t)W(x_w\rightarrow x_w+\delta x_w)\nonumber \\
  &=& \sum_{\delta x_w} P(x_w,t)W(x_w\rightarrow x_w+\delta x_w) - \frac{\partial}{\partial x_w}\sum_{\delta x_w}[\delta x_w P(x_w,t)W(x_w \rightarrow x_w +\delta x_w)]\nonumber \\ &+& \frac{1}{2}\frac{\partial^2}{\partial x_w^2}\sum_{\delta x_w}[\delta x_w^2 P(x_w,t)W(x_w\rightarrow x_w+\delta x_w)] - \sum_{\delta x_w} P(x_w,t)W(x_w\rightarrow x_w+\delta x_w)\nonumber \\
  &=& - \frac{\partial}{\partial y}[\overline A(y)P(y,t)] + \frac{1}{2}\frac{\partial^2}{\partial y^2}[\overline B(y)P(y,t)].\label{dw-fp1})
 \end{eqnarray}

Here $y\equiv x_w$ and
\begin{eqnarray}
 \overline A(x_w)&=& \sum_{\delta x_w}\delta x_w W(x_w \rightarrow x_w+\delta x_w)=\frac{1}{L\Delta(x_w)}[-q(0)\alpha(1-\alpha) + q(1) \beta(1-\beta)],\\
 \overline B(x_w)&=& \sum_{\delta x_w}\delta x_w^2 W(x_w \rightarrow x_w+\delta x_w)=\frac{1}{L^2 \Delta^2(x_w)}[q(0)\alpha(1-\alpha) + q(1) \beta(1-\beta)]>0,\nonumber \\ &&\;\;\;\;\;\;\;\;\;\;\;\;\;\;\;\;\;\;\;\;\;\;\;\;\;\;\;\;\;\;\;\;\;\;\;\;\;\;\;\;\;\;\;\;\;\;\;\;\;\;\;\;\;\;\;\;\;\;\;\;\;\;\;\;\;\;\;\;\;\;\;\;\text{for LD-HD transition},\label{Bxw}\\
 &=& \frac{1}{L^2 \Delta^2(x_w)}\left[\frac{q_\text{min}}{4} + q(0)\alpha(1-\alpha)\right]>0,\;\;\text{for LD-MC transition}, \label{Bxw-ld-mc} \\
 &=& \frac{1}{L^2 \Delta^2(x_w)}\left[q(1) \beta(1-\beta) + \frac{q_\text{min}}{4}\right]>0, \;\;\text{for HD-MC transition}. \label{Bxw-hd-mc}
\end{eqnarray}

\end{widetext}

Equation~(\ref{dw-fp1}) is the Fokker-Planck equation for the DW position $x_w$. The first term in the last line of (\ref{dw-fp1}) is the drift term, and the last term is the diffusion term.
At the delocalisation transition by using (\ref{deloc-cond}), we get $\overline A(x_w)=0$. Thus the drift term vanishes, giving
\begin{equation}
 \frac{\partial P}{\partial t}=\frac{\partial^2 }{\partial y^2}[\overline B(y)P(y)]=0,\label{dw-fp3}
\end{equation}
which is the Fokker-Planck equation at the delocalisation transition.  Treating $P(y)$ as a conserved diffusing density, Eq.~(\ref{dw-fp3}) is reminiscent of the conserved model B type relaxational dynamics~\cite{halperin} of a noninteracting density (an ideal gas), fluctuating about its mean, with an effective space-dependent inverse susceptibility or inverse compressibility $\chi^{-1}(y)\equiv \overline B(y)$ { away from any critical point}. Equation~(\ref{dw-fp3}) allows us to identify a ``current'' $j(y)=\frac{\partial}{\partial y}[\overline B(y)P(y)]$.

Now setting $\partial_t P=0$ in the steady state, we find
\begin{equation}
 \frac{\partial^2 }{\partial y^2}[\overline B(y)P(y)]=0,\label{dw-fp}
\end{equation}
at the delocalisation transition.  This has a general solution that $j(y)=j_0$, a constant. If $j_0$ is {\em non-zero}, this would imply a steady probability current in the configuration space defined by the TASEP lattice. This in turn should mean a steadily moving domain wall that will ultimately reach one end of the TASEP lane and leave it. This is unphysical, as a DDW randomly moves within the TASEP lane. We thus set $j_0=0$, giving $j(y)=0$. Thus the domain wall effectively executes equilibrium dynamics~\cite{tirtha-prr,akgupta}.  Thus, Eq.~(\ref{dw-fp3}) may be interpreted as the conserved model B type relaxational dynamics~\cite{halperin} of a noninteracting density an {\em equilibrium} ideal gas, a mapping that allows us to intuitively understand the our quantitative results; see below. By using $j(y)=0$,  we find
\begin{equation}
 P(y)=\frac{C}{\overline B(y)},\label{prob}
\end{equation}
where $C$ is a constant of integration. Recall that $y=x_w$, the position of the domain wall, and $\Delta(x_w)$ is the height of the domain wall, where
\begin{eqnarray}
 \Delta(x_w)= \rho_\text{HD}(x_w) - \rho_\text{LD}(x_w),\label{ld-hd-delta}\\
 \Delta(x_w)= \rho_\text{MC}(x_w) - \rho_\text{LD}(x_w),\label{ld-mc-delta}\\
 \Delta(x_w)= \rho_\text{HD}(x_w) - \rho_\text{MC}(x_w),\label{hd-mc-delta},
\end{eqnarray}
respectively, for a DDW at the LD-HD, LD-MC and HD-MC phase transitions. It is {\em space-dependent} here.  This is
in contrast to an open TASEP with uniform hopping rates, where $\Delta$ is a constant, independent of $x_w$. 
From \eqref{Bxw}, \eqref{Bxw-ld-mc}, \eqref{Bxw-hd-mc} and \eqref{prob}, we find the probability $P(x)$ is given by
\begin{eqnarray}
 &&P(x)\propto \overline B(x)^{-1}\propto [\rho_\text{HD}(x)-\rho_\text{LD}(x)]^2,\label{prob1}\\
 &&P(x)\propto \overline B(x)^{-1}\propto [\rho_\text{MC}(x)-\rho_\text{LD}(x)]^2,\label{prob-ld-mc}\\
 &&P(x)\propto \overline B(x)^{-1}\propto [\rho_\text{HD}(x)-\rho_\text{MC}(x)]^2,\label{prob-hd-mc}
\end{eqnarray}
for the LD-HD, LD-MC and HD-MC transitions, which
depending on the coordinate $x$ is maximum (minimum), where the height difference between the two coexistence stationary density solutions are maximum (minimum).  Using our conceptual mapping to an equilibrium ideal  gas with a nonuniform compressibility, we note that $P(x)$, a measure of the local density fluctuations in this intuitive picture in terms of a conserved diffusing density, rises in regions with {\em higher} compressibility (or lower $\overline B(x)$). This is easy to understand: in regions with higher compressibility, the free energy cost of higher fluctuations is smaller, letting the gas fluctuate more. In the domain wall picture (as here), a larger $P(x)$ in a region implies, larger probability of finding the DW in that region. Notice that in the equivalent picture in terms of an ideal gas, its compressibility $\propto \Delta^2$ locally. 

We now calculate the density profile of the DDW $\rho_\text{DDW}(x)$ starting by relating it to $P(x)$. As argued in Ref.~\cite{tirtha-prr}, we must have
\begin{equation}
 \frac{d\rho_\text{DDW}}{dx}=DP(x),\label{basic-rho}
\end{equation}
where $D>0$ is a constant. Solving (\ref{basic-rho}), we get
\begin{eqnarray}
 \rho_\text{DDW}(x)&=&D\int\,dx\, P(x) + B 
 \nonumber \\&=& A\int\,dx\,[\rho_\text{HD}(x)-\rho_\text{LD}(x)]^2 +B.\label{rho-ddw-gen}
\end{eqnarray}
Constants $A$ and $B$ may be fixed by using the boundary conditions.


 
 
 Using the above theory, we first study the envelop of a DDW at the LD-HD phase coexistence, and then the same for LD-MC and HD-MC phase coexistences.

 \subsection{DDW envelops at the LD-HD phase coexistence}

In this Section, we calculate the DDW envelop at the LD-HD phase coexistence. { Here, $J_\text{LD}= J_\text{in}$ and $J_\text{HD}=J_\text{out}$ are physically the ``injection limited'' and ``extraction limited'' stationary currents.}
Notice that the LD-HD coexistence line takes the usual form $\alpha=\beta$ (as in an open TASEP), so long as $q(0)=q(1)$, {\em even if} $q(x)$ in general is {\em nonuniform}. 
In an open uniform TASEP, a DDW profile connects $\rho_\text{LD}=\alpha$ at $x=0$ and $\rho_\text{HD}=1-\beta$ at $x=1$ by an inclined straight line with parameters $\alpha=\beta$, which is the coexistence line in the $\alpha-\beta$ plane. In the present study, with space dependent $q(x)$, the DDW profile still connects $\rho_\text{LD}(x=0)=\alpha$ and $\rho_\text{HD}(x=1)=1-\beta$ (however $\alpha\neq \beta$ in general), but {\em not} through a straight line anymore. Our aim in this Section is to calculate the equation for this line by using the above theoretical framework. 

 We first set the boundary conditions to calculate the constants $A$ and $B$ in \eqref{rho-ddw-gen}. In the LD-HD phase coexistence, the boundary conditions are $\rho_\text{DDW}(0)=\alpha,\,\rho_\text{DDW}(1)=1-\beta$. Now, use \eqref{rho-ld} and \eqref{rho-hd} respectively for $\rho_\text{LD}(x)$  and $\rho_\text{HD}(x)$. For a DDW,
\begin{equation}
 J_\text{LD}=q(0)\alpha(1-\alpha)=J_\text{HD}=q(1)\beta(1-\beta)=J_\text{DDW}.
\end{equation}
We then find
\begin{eqnarray}
 \Delta (x)= \rho_\text{HD}(x)-\rho_\text{LD}(x)=\sqrt{1-\frac{4J_\text{DDW}}{q(x)}},\label{ddw-height}
\end{eqnarray}
which is the domain wall ``height'' at $x$. 
 
 We now apply the above theory to several specific examples delineated by different forms for the hopping rate function $q(x)$ and calculate the shapes of the DDWs on the respective coexistence lines.  Before we consider various choices for $q(x)$, we consider the relation between the analytic structure of $q(x)$ and the continuity properties of probability $P(x)$ and by extension of $\rho_\text{DDW}(x)$, which will be useful to determine the constants of integration in (\ref{rho-ddw-gen}). By using (\ref{prob1}) and the expressions (\ref{rho-ld}) and (\ref{rho-hd}) above, we note that $P(x)$ is discontinuous only where $q(x)$ is discontinuous. Since $q(x)$ only has finite discontinuities at isolated points and is otherwise smoothly varying, so does $P(x)$. From (\ref{basic-rho}), $d\rho_\text{DDW}(x)/dx$ that scales with $P(x)$ must have finite discontinuities at those points only, elsewhere, it must be continuous. It then immediately follows that $\rho_\text{DDW}(x)$ must be continuous {\em everywhere}, even at points where $q(x)$ and hence $P(x)$ are {\em discontinuous}. The latter discontinuities are reflected in $\rho_\text{DDW}(x)$ as ``kinks''. These are important general predictions from our analysis. Our choices for $q(x)$ below to study the DDW envelops can be used to confirm these general expectations.  \\

\begin{figure}
 \includegraphics[width=8.7cm]{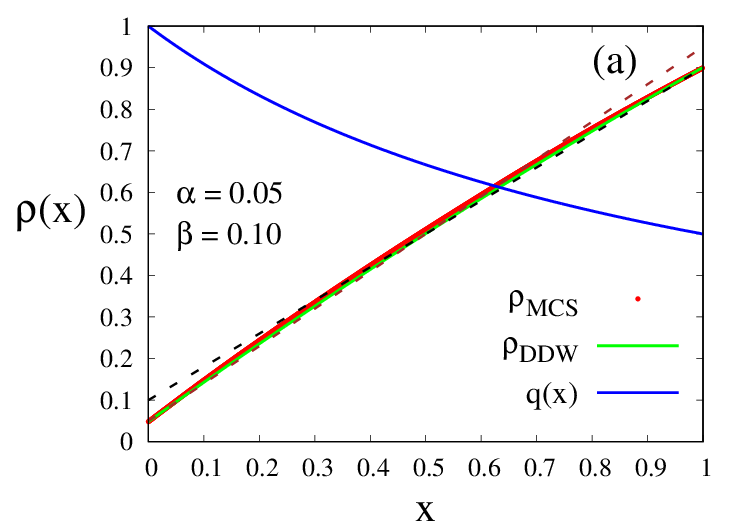}\\ \includegraphics[width=4.3cm]{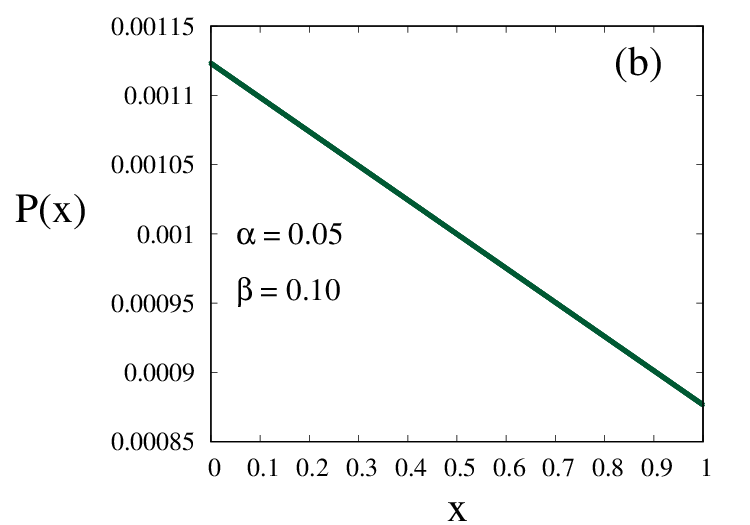}\hfill \includegraphics[width=4.3cm]{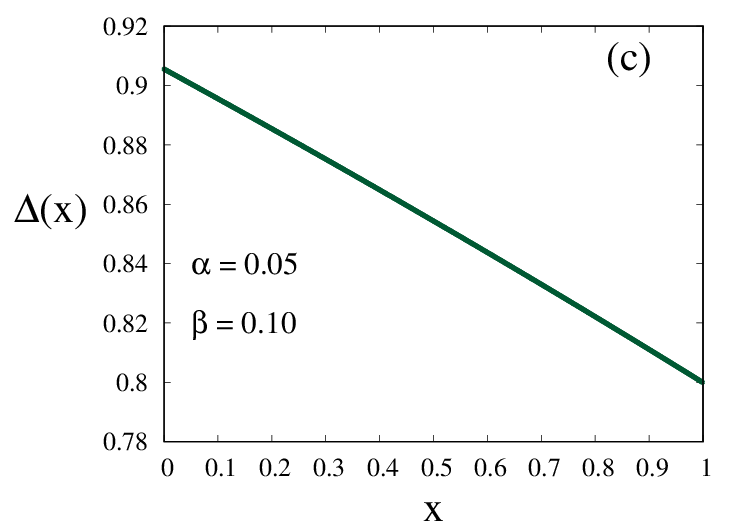}
\caption{(a) Analytically obtained DDW envelope $\rho_\text{DDW}(x)$ in (\ref{ddw1}) for $\alpha,\beta$ on the coexistence line (\ref{coex-1}) with $q(x)$ as given in (\ref{ddw-q1}) and $\alpha=0.05,\beta=0.1$. Both analytical result  (green line) and MCS (red dots, system size $L=1000$) are shown, which agree well with each other. The broken lines are reference lines, which would have been the DDW profiles with $\alpha=\beta=0.05$ and $\alpha=\beta=0.1$ in an open, uniform TASEP. The blue curve denotes the hopping rate function $q(x)$. Plots of the (b) corresponding local probability $P(x)$  and (c) domain wall height $\Delta(x)$  are shown. Both are space-dependent; see text.    }
\label{ddw-fig1}
\end{figure}

 (i) We first consider 
 \begin{equation}
  q(x)=\frac{1}{1+x},\label{ddw-q1}
 \end{equation}
such that at $x=0$, $q(0)=1$ (maximum) and at $x=1$, $q(1)=1/2$ (minimum). The equation for the coexistence line in the present case is obtained from (\ref{coex}) and reads
\begin{equation}
 \beta=\frac{1}{2}\left[1-\sqrt{1-8\alpha(1-\alpha)}\right].\label{coex-1}
\end{equation}
By using \eqref{rho-ddw-gen}, we then get the stationary density on the coexistence line (\ref{coex-1}):
 \begin{equation}
  \rho_\text{DDW}(x)=Ax(1-4J_\text{DDW})-2AJ_\text{DDW}x^2 +B.
 \end{equation}
Using the boundary conditions $\rho_\text{DDW}(x=0)=\alpha$ and $\rho_\text{DDW}(x=1)=1-\beta$, we find
\begin{equation}
 A=\frac{1-\beta-\alpha}{1-6J_\text{DDW}},\,B=\alpha,
\end{equation}
giving
\begin{equation}
 \rho_\text{DDW}(x)=\frac{1-\beta-\alpha}{1-6J_\text{DDW}}[x(1-4J_\text{DDW})-2J_\text{DDW}x^2] +\alpha.\label{ddw1}
\end{equation}
The analytical DDW profile $\rho_\text{DDW}(x)$ as given in (\ref{ddw1}) has been plotted (green line) in Fig.~\ref{ddw-fig1}(a) together with the corresponding MCS result (red dots). Good agreements between the two is visible. We have also plotted $P(x)$ and $\Delta(x)$ from our analytical treatment; see Fig.~\ref{ddw-fig1}(b) and Fig.~\ref{ddw-fig1}(c) respectively. 


\vskip0.3cm

\begin{figure}[htb]
 \includegraphics[width=8.7cm]{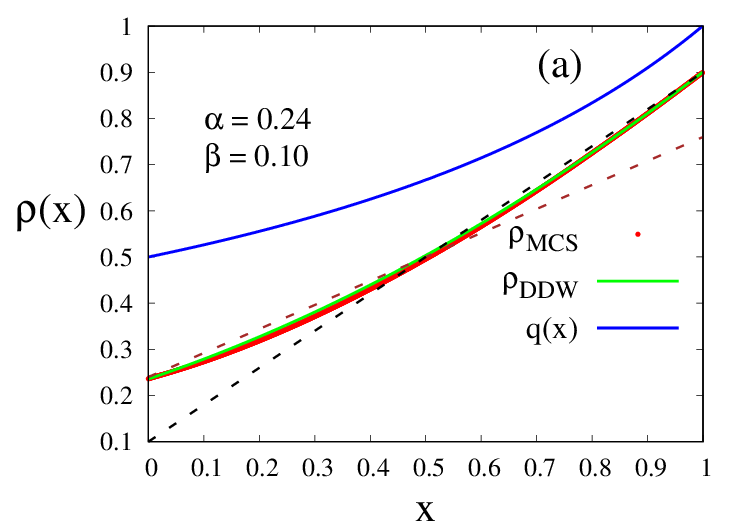}\\ \includegraphics[width=4.2cm]{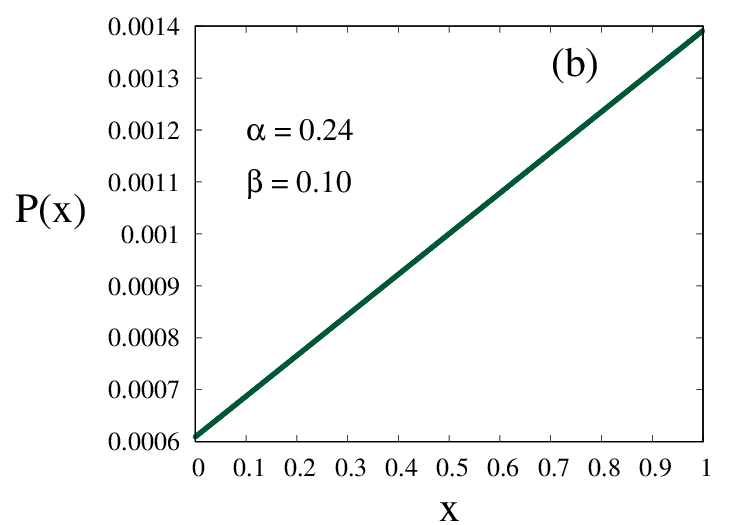}\hfill
 \includegraphics[width=4.3cm]{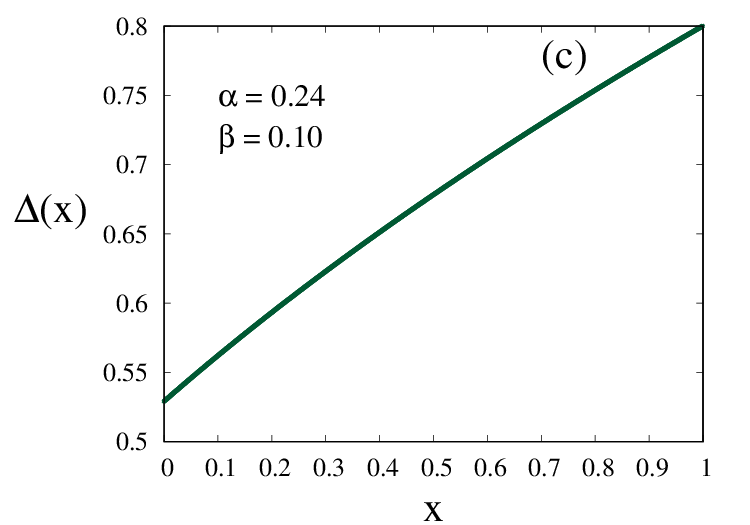}
  \caption{(a) Analytical result on DDW envelope (\ref{ddw2}) for $\rho_\text{DDW}(x)$ for $\alpha,\beta$ on the coexistence line (\ref{coex-2}) with $q(x)$ as given in (\ref{ddw-q2}) and $\alpha=0.24,\beta=0.1$. Both analytical density  (green line) and MCS (red dots, system size $L=1000$) are shown, which agree well with each other. The broken lines are reference lines, which would have been the DDW profiles with $\alpha=\beta=0.24$ and $\alpha=\beta=0.1$ in an open, uniform TASEP. The blue
curve denotes the hopping rate function $q(x)$. Plots of the (b) corresponding local probability $P(x)$  and (c) domain wall height $\Delta(x)$  are shown. Both are space-dependent. See text. }
  \label{ddw-fig2}
 \end{figure}

(ii) Next, we consider 
\begin{equation}
q(x)=\frac{1}{2-x}.\label{ddw-q2}
\end{equation}
Thus, at $x=0$, $q(0)=1/2$ (minimum) and at $x=1$, $q(1)=1$ (maximum). The equation for the coexistence line in the present case is obtained from (\ref{coex}) and reads
\begin{equation}
\alpha=\frac{1}{2}\left[1-\sqrt{1-8\alpha(1-\alpha)}\right]. \label{coex-2}
\end{equation}
By using the boundary conditions $\rho_\text{DDW}(x=0)=\alpha$ and $\rho_\text{DDW}(x=1)=1-\beta$, we get
\begin{equation}
 \rho_\text{DDW}(x)=\frac{1-\alpha-\beta}{1-6J_\text{DDW}}[x(1-8J_\text{DDW})+2J_\text{DDW}x^2]+\alpha.\label{ddw2}
\end{equation}
We have presented the above DDW profile (\ref{ddw2}) along with the corresponding MCS results in Fig.~\ref{ddw-fig2}(a), showing good agreements between the two. We have also plotted $P(x)$ or $\Delta(x)$ from our analytical treatment; see Fig.~\ref{ddw-fig2}(b) and Fig.~\ref{ddw-fig2}(c). 

 
\vskip0.3cm

(iii) So far we have studied the DDW profiles with $q(x)$, which are {\em continuous}, and also {\em smooth}, i.e., $dq(x)/dx$ is also continuous. We now consider a case, where $q(x)$ is continuous, but not smooth, i.e., $dq(x)/dx$ has a finite discontinuity in the bulk. For that, we consider a hopping rate function
\begin{eqnarray}
 q(x)&=& \frac{1}{1+ax},\;\;\;0\leq x\leq x_0,\nonumber \\
     &=& \frac{1}{3-bx},\;\;\;x_0\leq x\leq 1. \label{gen-qx}
\end{eqnarray}
Here, $a,b$ are positive constants. What is $x_0$? It is determined by the imposed condition $q(x\rightarrow x_{0+}) = q(x\rightarrow x_{0-})$; { $x_0$ is also the location of the minimum of $q(x)$ here}. This gives
\begin{equation}
 x_0=\frac{2}{a+b}.\label{x0}
\end{equation}
Given the form of $q(x)$ in (\ref{gen-qx}), we find that if $x_0=1/2$ along with $a=b=2$, $q(x)$ is symmetric about $x=1/2$, the midpoint of the system; else, for any $x_0\neq 1/2$, it is asymmetric.
Clearly, $dq/dx$ has a finite discontinuity at $x=x_0$. Since \eqref{basic-rho} is valid in the full domain of $x\in [0,1]$, we write the general solution to \eqref{basic-rho} for $q(x)$ as in (\ref{gen-qx}) as
\begin{eqnarray}
 \rho_\text{DDW}(x)&=& A\int\,dx\,[\rho_\text{HD}(x)-\rho_\text{LD}(x)]^2 +B_1,\,\,0\leq x\leq x_0\nonumber \\
                   &=&  A\int\,dx\,[\rho_\text{HD}(x)-\rho_\text{LD}(x)]^2 +B_2,\,\,x_0\leq x\leq 1.\nonumber \\\label{gen-rho1}
\end{eqnarray}
The above form in (\ref{gen-rho1}) for $\rho_\text{DDW}(x)$  automatically ensures that $d\rho_\text{DDW}/dx$ follows the same equation in the entire domain $x\in [0,1]$. Since $q(x)$ is continuous everywhere, $\rho_\text{DDW}(x)$ must also be continuous everywhere.  Constants $A,B_1,B_2$ may be evaluated from the boundary conditions on $\rho_\text{DDW}(0), \rho_\text{DDW}(1)$ and the condition $\rho_\text{DDW}(x\rightarrow x_{0-}) = \rho_\text{DDW}(x\rightarrow x_{0+})$. We now use (\ref{gen-qx}) to obtain the corresponding $\rho_\text{DDW}(x)$, which generally reads
\begin{subequations}
\begin{eqnarray}
\rho_\text{DDW}(x)&=&A(1-4 J_\text{DDW})x - 2AJ_\text{DDW} ax^2 + B_1,\nonumber\\
&&\;\;\;\;\;\;0\leq x\leq x_0, \\
                  &=& A (1-12 J_\text{DDW})x + 2A J_\text{DDW}bx^2 + B_2,\nonumber\\
                  &&\;\;\;\;\;\;x_0\leq x\leq 1.
\end{eqnarray}
\label{rho44}
\end{subequations}
Solving for the constants with the boundary conditions specified above, we find
\begin{subequations}
\begin{eqnarray}
 A&=& \frac{1-\alpha-\beta}{J_\text{DDW}[8x_0 -2x_0^2 (a+b) -12 +2b]+1}, \label{Ano}\\
 B_1&=& \alpha, \label{B1no}\\ 
 B_2&=&1-\beta - A(1-12 J_\text{DDW}+2J_\text{DDW}b).\label{gen-rho-ab}
\end{eqnarray}
\end{subequations}

\begin{figure}
  \includegraphics[width=8.7cm]{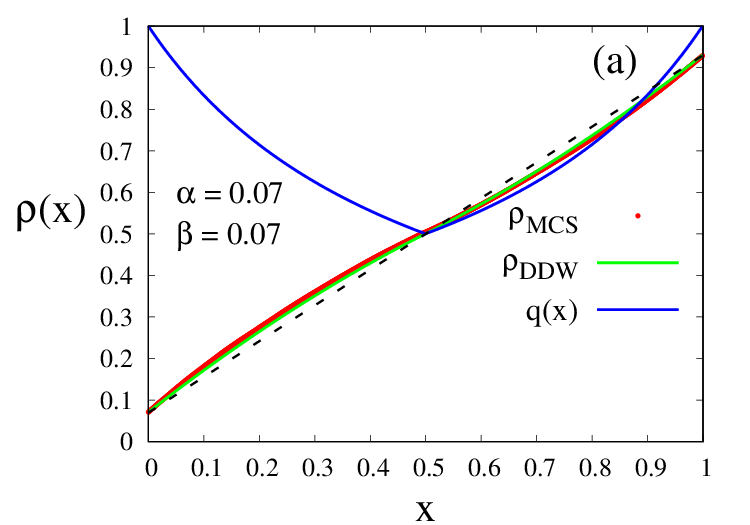}\\  \includegraphics[width=4.3cm]{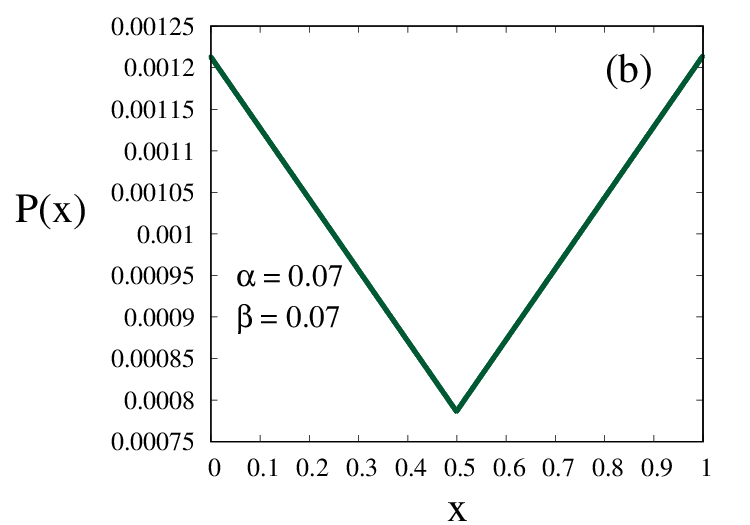}\hfill \includegraphics[width=4.3cm]{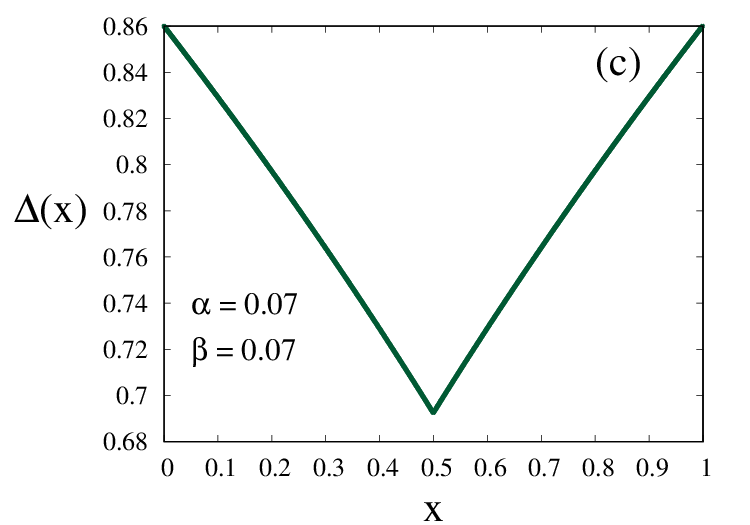}
  \caption{
  (a) Analytically obtained DDW profile $\rho_\text{DDW}(x)$  (\ref{rho44}) with (\ref{Ano}), (\ref{B1no}) and (\ref{gen-rho-ab}) for $\alpha,\beta$ on the coexistence line (\ref{coex-3}) with $q(x)$ as given in (\ref{discdq-sym}) and $\alpha=0.07,\beta=0.07$. Both analytical result (green line) and MCS (red dots, system size $L=1000$) are shown, which agree well with each other. The broken line is a reference line, which would have been the DDW envelope with $\alpha=0.07,\beta=0.07$ in an open, uniform TASEP. The deviation of $\rho_\text{DDW}(x)$ from the DDW in an open, uniform TASEP is clearly visible. The blue
curve denotes the hopping rate function $q(x)$. Plots of the corresponding (b) local probability $P(x)$  and (c) domain wall height $\Delta(x)$ from our analytical treatment are shown. Both are space-dependent. See text. }
  \label{ddw-fig3}
   \end{figure}

We study the two following cases:

(a) $x_0=1/2$. We choose $a=b=2$, giving 
\begin{eqnarray}
 q(x)&=& \frac{1}{1+2x},\;\;0\leq x\leq \frac{1}{2},\nonumber\\
     &=& \frac{1}{3-2x},\;\;\frac{1}{2}\leq x\leq 1.\label{discdq-sym}
\end{eqnarray}
The equation for the coexistence line in the present case is obtained from (\ref{coex}) and reads
\begin{equation}
 \alpha=\beta, \label{coex-3}
\end{equation}
identical to that for a uniform open TASEP.
Analytical result for $\rho_\text{DDW}(x)$ is then obtained from (\ref{rho44}), which is shown in Fig.~\ref{ddw-fig3}(a). The corresponding MCS results are also plotted, which agree well with the analytical result. However, both the MFT and MCS results for the stationary density clearly deviates from its counterpart in an open, uniform TASEP with $\alpha=0.07,\beta=0.07$. This deviation is due to the space dependence of $q(x)$. The local normalised probability distribution $P(x)$ [see Eq.~(\ref{prob})] and domain wall height $\Delta(x)$ [see Eq.~(\ref{ddw-height})] in this case are shown, respectively, in Fig.~\ref{ddw-fig3}(b) and Fig.~\ref{ddw-fig3}(c). Both are space dependent.

 
(b) Next we choose $x_0=1/3$. We choose $a=4, b=2$, giving 
\begin{eqnarray}
 q(x)&=& \frac{1}{1+4x},\;\;0\leq x\leq \frac{1}{3},\nonumber\\
     &=& \frac{1}{3-2x},\;\;\frac{1}{3}\leq x\leq 1.\label{discdq-asym}
\end{eqnarray}
The equation for the coexistence line in the present case is obtained from (\ref{coex}) and reads
\begin{equation}
 \alpha=\beta, \label{coex-4}
\end{equation}
identical to that in a uniform open TASEP. Notice that the coexistence line (\ref{coex-4}) is identical to the previous case (\ref{coex-3}), {\em in spite of} having different $q(x)$ in these two cases. This is because, the coexistence line is controlled by the ratio of $q(0)$ and $q(1)$, which is unity in both these cases.

\begin{figure}
  \includegraphics[width=8.7cm]{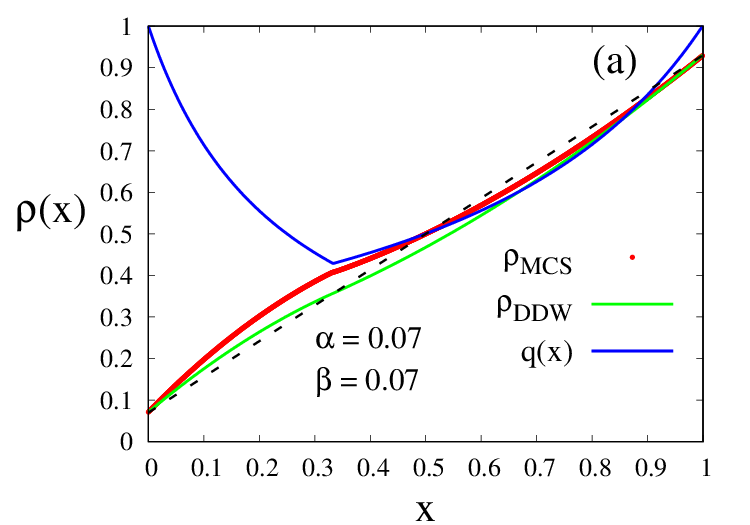}\\ \includegraphics[width=4.3cm]{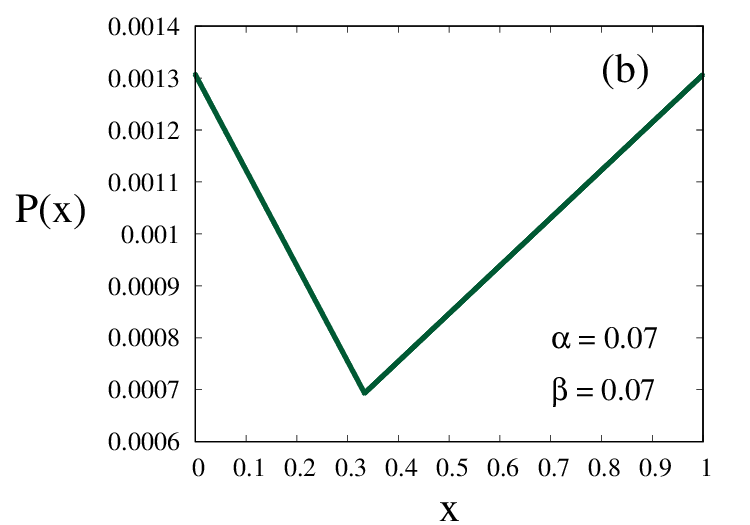}\hfill
  \includegraphics[width=4.3cm]{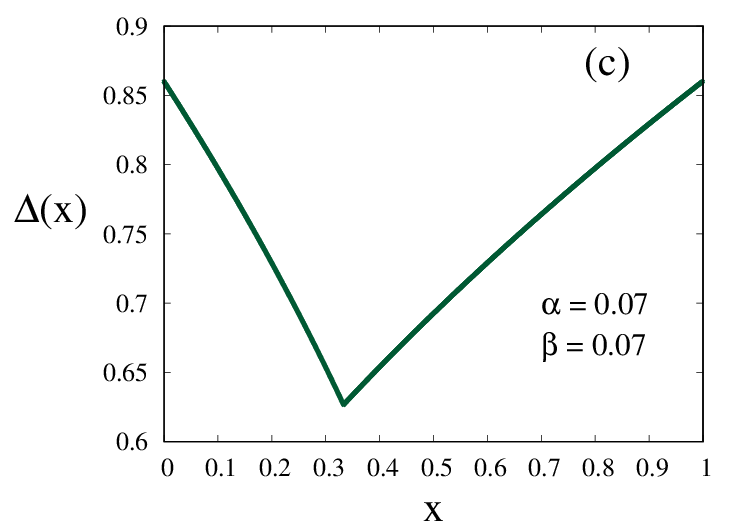}
  \caption{(a) Analytically obtained DDW profile $\rho_\text{DDW}(x)$  (\ref{rho44}) with (\ref{Ano}), (\ref{B1no}) and (\ref{gen-rho-ab}) for $\alpha,\beta$ on the coexistence line (\ref{coex-4}) with $q(x)$ as given in (\ref{discdq-asym}) and $\alpha=0.07,\beta=0.07$. Both analytical result (green line) and MCS (red dots, system size $L=1000$) are shown, which agree well with each other.  The broken line is a reference line, which would have been the DDW envelope with $\alpha=\beta=0.07$ in an open, uniform TASEP. The blue
curve denotes the hopping rate function $q(x)$. Plots of the corresponding (b) local probability $P(x)$  and (c) domain wall height $\Delta(x)$  are shown. Both are space-dependent; see text.}
  \label{ddw-fig4}
 \end{figure}

We obtain the analytical result for $\rho_\text{DDW}(x)$ from (\ref{Ano})-(\ref{gen-rho-ab}). This is shown in Fig.~\ref{ddw-fig4}(a) in green lines. The corresponding MCS results (red dots) are also plotted in Fig.~\ref{ddw-fig4}(a).  The MFT and MCS results generally agree  with each other. Both however clearly deviate from  an open, uniform TASEP with $\alpha=0.07,\beta=0.07$. This deviation is due to the space dependence of $q(x)$. The local normalised probability distribution $P(x)$ [see Eq.~(\ref{prob})] and domain wall height $\Delta(x)$ [see Eq.~(\ref{ddw-height})] in this case are shown in Fig.~\ref{ddw-fig4}(b) and Fig.~\ref{ddw-fig4}(c) respectively. Both are space dependent.

 
\vskip0.3cm

(iv)
Now we study the stationary densities with hopping rate functions $q(x)$ having a finite discontinuity at $x=x_0$, in contrast to the previous cases where $q(x)$ is assumed to be continuous. As explained earlier, since the discontinuity of $q(x)$ is {\em finite}, from (\ref{rho-ddw-gen}) $\rho_\text{DDW}(x)$ must also be continuous at $x=x_0$, the point of discontinuity of $q(x)$. Nonetheless, $d\rho_\text{DDW}/dx$ and hence probability $P(x)$ of finding the domain wall at $x$ are {\em discontinuous} at $x=x_0$.  Thus, $\rho_\text{DDW}(x)$ should have a kink at $x=x_0$. We study two distinct examples below.

In the first case we use $q(x)$ as given in (\ref{case2}) above. In this case,
such that $q(x)$ is discontinuous at $x=1/2$. The discontinuity is finite of size 1/2. 
The equation for the coexistence line can be obtained from (\ref{coex}) and is given by (\ref{coex-1}), since $q(0)=1,\,q(1)=1/2$, same as for case (i) above. We thus see that the discontinuity in $q(x)$ has no effect on the equation of the coexistence line. However, the DDW profile $\rho_\text{DDW}(x)$ has a kink at the location of the discontinuity of $q(x)$.

\begin{figure}[htb]
  \includegraphics[width=8.7cm]{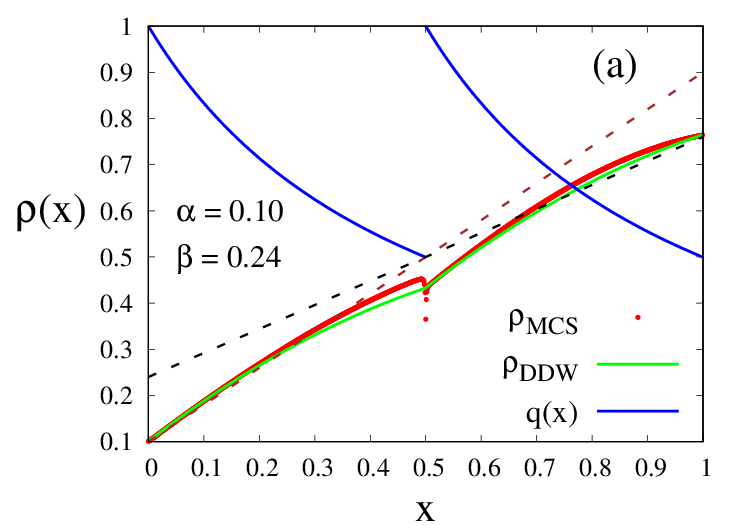}\\ \includegraphics[width=4.3cm]{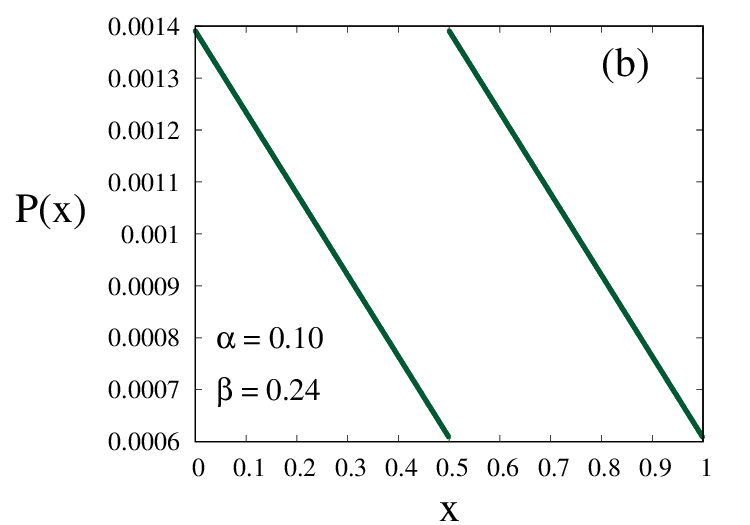}\hfill 
  \includegraphics[width=4.3cm]{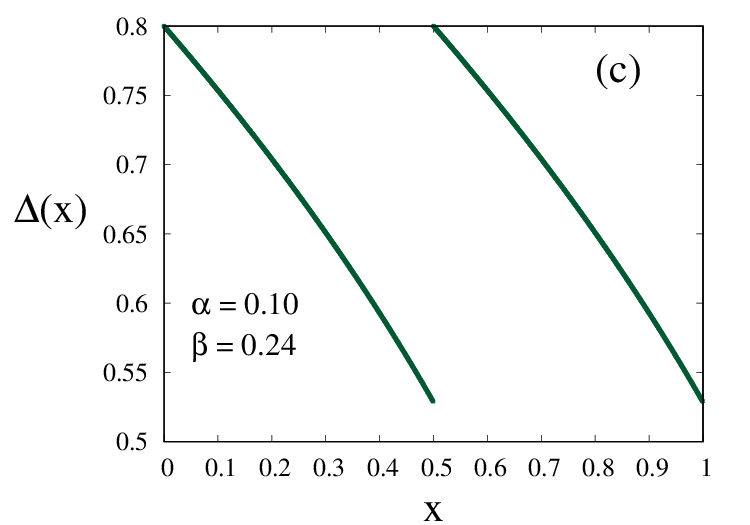}
  \caption{(a) Analytically obtained DDW profile $\rho_\text{DDW}(x)$ in (\ref{rho50})  with (\ref{ddw5B1}), (\ref{ddw5B2}) and (\ref{ddw-5}) for $\alpha,\beta$ on the coexistence line (\ref{coex-1}) with $q(x)$ as given in (\ref{case2}) and $\alpha=0.1,\beta=0.24$. Both analytical result (green line) and MCS (red dots, system size $L=1000$) are shown, which agree well with each other. The broken lines correspond to reference  lines, which would have been the DDW profiles in an open, uniform TASEP with $\alpha=\beta=0.1$ and $\alpha=\beta=0.24$.  Plots of the corresponding (b) local probability $P(x)$  and (c) domain wall height $\Delta(x)$  are shown. Both are space-dependent. See text.}
  \label{ddw-fig5}
  \end{figure}

We continue to use the general solution for $\rho_\text{DDW}(x)$ as given in \eqref{gen-rho1}.  We find
\begin{subequations}
\begin{eqnarray}
 \rho_\text{DDW}(x)&=&A[x-4J_\text{DDW}x-4J_\text{DDW}x^2]+B_1,\nonumber \\
 &&0\leq x\leq 1/2, \\
                   &=&A[x-4J_\text{DDW}x^2] + B_2,\,1/2\leq x\leq 1.
\end{eqnarray}
\label{rho50}
\end{subequations}
The coefficients $A,\,B_1\,B_2$ are to be evaluated by using the boundary conditions (i) $\rho_\text{DDW}(x=0)=\alpha$, (ii)$\rho_\text{DDW}(x=1)=1-\beta$ and (iii) $\rho_\text{DDW}(x\rightarrow 1/2_-)=\rho_\text{DDW}(x\rightarrow 1/2_+)$.
We find
\begin{subequations}
\begin{eqnarray}
 B_1&=& \alpha,\label{ddw5B1}\\
 B_2&=& \frac{1}{1-6J_\text{DDW}}\bigg[\alpha - 4\alpha J_\text{DDW}\nonumber \\&-&2J_\text{DDW}+ 2J_\text{DDW}\beta\bigg],\label{ddw5B2} \\
 A&=& \frac{1-\beta-B_2}{1-4J_\text{DDW}},\label{ddw-5}
\end{eqnarray}
\end{subequations}
giving us $\rho_\text{DDW}(x)$ for a hopping rate function (\ref{case2}). We have shown plots of $\rho_\text{DDW}(x)$ versus $x$ from our analytical treatment, superposed with the corresponding MCS result, in Fig.~\ref{ddw-fig5}(a). As expected and similar to the previous cases, $\rho_\text{DDW}(x)$ is continuous but now has a kink at  $x=1/2$, the location of the discontinuity of $q(x)$.  Once again similar to the previous cases, $\rho_\text{DDW}(x)$ clearly deviates from its counterparts in an open, uniform TASEP either with $\alpha=0.1,\beta=0.1$, or with $\alpha=0.24,\beta=0.24$. This deviation is due to the space dependence of $q(x)$. The local normalised probability distribution $P(x)$ [see Eq.~(\ref{prob})] and domain wall height $\Delta(x)$ [see Eq.~(\ref{ddw-height})] in this case are shown in Fig.~\ref{ddw-fig5}(b) and Fig.~\ref{ddw-fig5}(c), respectively. Unsurprisingly, both are space dependent. The discontinuity of $q(x)$ is clearly visible in the discontinuities of $P(x)$ and $\Delta(x)$.

Finally, we study  a DDW with $q(x)$ as given in (\ref{case3}) above.
The equation for the coexistence line can be obtained from (\ref{coex}) and is given by (\ref{coex-2}), since $q(0)=1/2,\,q(1)=1$, same as for case (ii) above. Nonetheless, $\rho_\text{DDW}(x)$ is affected by the specific for of $q(x)$ in (\ref{case3}) as we show below.

\begin{figure}[htb]
 \includegraphics[width=8.7cm]{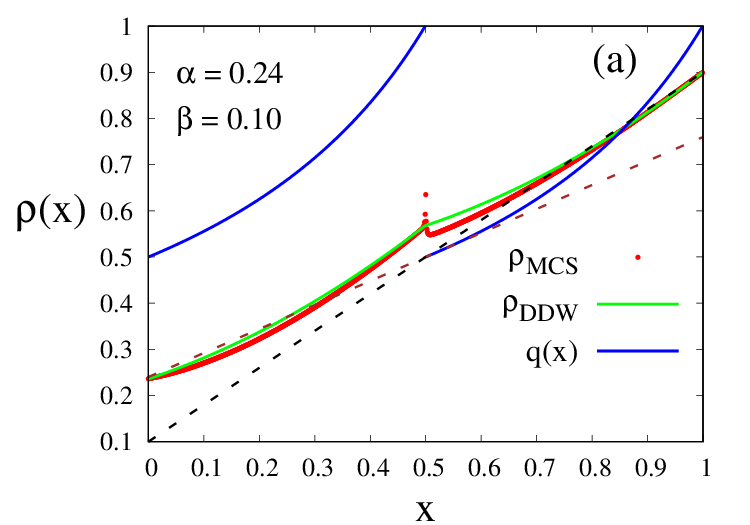}\\ \includegraphics[width=4.3cm]{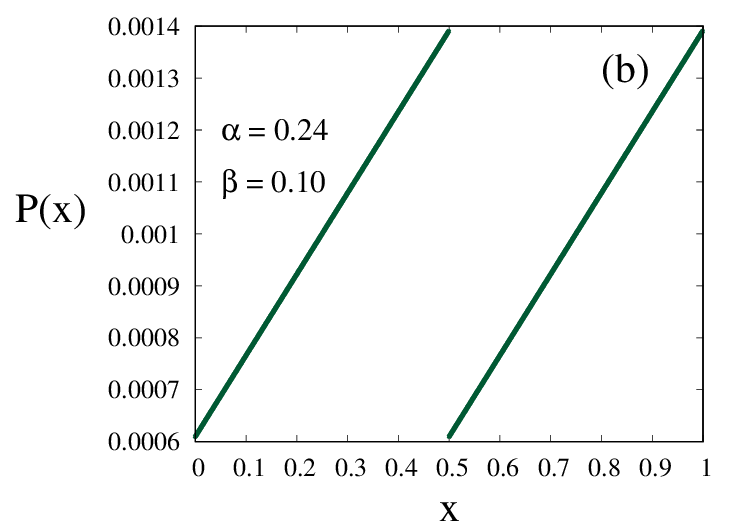}\hfill 
 \includegraphics[width=4.3cm]{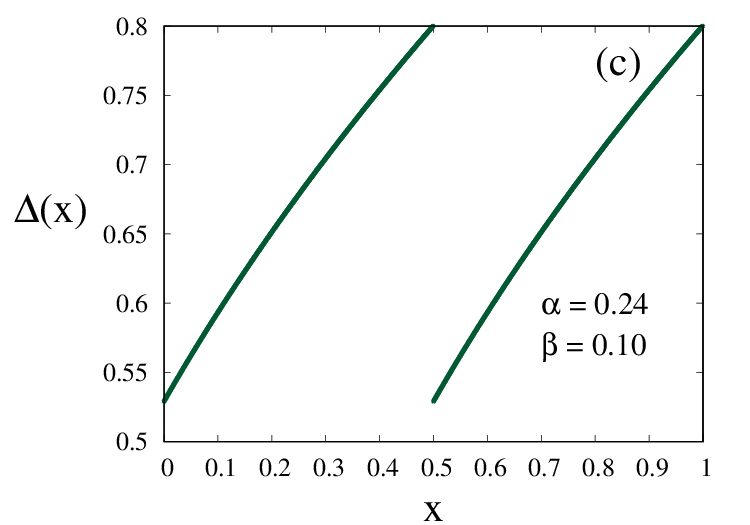}
 \caption{(a) Analytically obtained density profile $\rho_\text{DDW}(x)$ in (\ref{rho52}) with (\ref{ddw6B1}), (\ref{ddw6B2}) and (\ref{ddw-6}) for $\alpha,\beta$ on the coexistence line (\ref{coex-2}) with $q(x)$ as given in (\ref{case3}) and $\alpha=0.1,\beta=0.24$. Both analytical result (green line) and MCS (red dots, system size $L=1000$) are shown, which agree well with each other. The broken lines are reference lines, which would have been the DDW profiles in an open, uniform TASEP with $\alpha=\beta=0.1$ and $\alpha=\beta=0.24$. The blue
curve denotes the hopping rate function $q(x)$. Plots of the corresponding (b) local probability $P(x)$  and (c) domain wall height $\Delta(x)$ are shown. Both are space-dependent. See text.}
 \label{ddw-fig6}
 \end{figure}

Using (\ref{gen-rho1}) together with (\ref{case3}), we get
\begin{subequations}
\begin{eqnarray}
 \rho_\text{DDW}(x)&=&A[x-8J_\text{DDW}x+4J_\text{DDW}x^2]+B_1,\nonumber \\&&\;\;\;\;\;\;\;\;\;\,0\leq x\leq \frac{1}{2}, \\
 &=& A[x-12 J_\text{DDW}x+4J_\text{DDW}x^2]+B_2,\nonumber \\&&\;\;\;\;\;\;\;\;\;\;\,\frac{1}{2}\leq x\leq 1.
\end{eqnarray}
\label{rho52}
\end{subequations}
As before, the constants $A,B_1,B_2$ are to be evaluated by using the boundary conditions (i) $\rho_\text{DDW}(x=0)=\alpha$, (ii)$\rho_\text{DDW}(x=1)=1-\beta$ and (iii) $\rho_\text{DDW}(x\rightarrow 1/2_-)=\rho_\text{DDW}(x\rightarrow 1/2_+)$. We find
\begin{subequations}
\begin{eqnarray}
 B_1&=&\alpha,\label{ddw6B1} \\
 B_2&=&\frac{2(1-\beta)J_\text{DDW}+\alpha(1-8J_\text{DDW})}{1-6J_\text{DDW}},\label{ddw6B2} \\
 A&=& \frac{1-\beta-B_2}{1-8J_\text{DDW}},\label{ddw-6}
 \end{eqnarray}
 \end{subequations}
giving us $\rho_\text{DDW}(x)$ for a hopping rate function given in (\ref{case3}). We have shown plots of $\rho_\text{DDW}(x)$ versus $x$ from our analytical treatment, superposed with the corresponding MCS result, in Fig.~\ref{ddw-fig6}(a). As expected and similar to the previous cases, $\rho_\text{DDW}(x)$ remains continuous but has a kink at  $x=1/2$, the location of the discontinuity of $q(x)$.  Unsurprisingly, $\rho_\text{DDW}(x)$ clearly deviates from its counterparts in an open, uniform TASEP either with $\alpha=0.1,\beta=0.1$, or with $\alpha=0.24,\beta=0.24$. This deviation is due to the space dependence of $q(x)$. The local normalised probability distribution $P(x)$ [see Eq.~(\ref{prob})] and domain wall height $\Delta(x)$ [see Eq.~(\ref{ddw-height})] in this case as obtained from our analytical treatment are shown in Fig.~\ref{ddw-fig6}(a) and Fig.~\ref{ddw-fig6}(b), respectively, which display discontinuity at $x=1/2$, a reflection of the discontinuity of $q(x)$ at that point. Both are space dependent.


{
    
\subsection{DDW envelops at the LD-MC and HD-MC phase coexistence} 
 
 Having analysed the DDW envelops at the LD-HD phase transitions, we now calculate the DDW envelops at the LD-MC and HD-MC transitions, {\em when} those transitions are discontinuous. While we will use the calculational logic outlined in the above Section for DDW envelops at the LD-HD phase transitions, we will see below that additional complications and novel features, {\em not} possible for a DDW at the LD-HD transition, can now arise. These complications have their origin in the fact that unlike for an open, uniform TASEP, the stationary density profile $\rho_\text{MC}(x)$ in the MC phase is a combination of $\rho_\text{LD}(x)$, the LD phase stationary density, and $\rho_\text{HD}(x)$, the HD phase stationary density. Consider \eqref{prob-ld-mc}, which gives the probability $P(x)$ of finding a DDW in $x$ at the LD-MC phase transition. Now if $\rho_\text{MC}(x)$ consists of $\rho_\text{HD}(x)$ over some range of $x$ and $\rho_\text{LD}(x)$ in the remaining range of $x$, $P(x)$ vanishes where $\rho_\text{MC}(x)=\rho_\text{LD}(x)$ and is nonzero only in the range of $x$ where $\rho_\text{MC}(x)=\rho_\text{HD}(x)$.  This in turn means that at the LD-MC transition {\em no} DDW is to be observed in the range of $x$, where $\rho_\text{MC}(x)=\rho_\text{LD}(x)$, but a DDW should be  observed only in the range of $x$ where $\rho_\text{MC}(x)=\rho_\text{HD}(x)$. Therefore, the DDW is {\em partial}, i.e., covers the TASEP lane {\em only partially}. Similarly, from \eqref{prob-hd-mc} we can conclude that a partial DDW should be found at the HD-MC phase transition for a range of $x$ in which $\rho_\text{MC}(x)=\rho_\text{LD}(x)$. Our detailed MCS studies, which are presented below, validate these arguments.
 
 We first calculate the DDW envelops at the LD-MC transition for a few specific choices of $q(x)$.
 
 (i) $q(x)=\frac{1}{1+x}$, which is (\ref{qx-ddw1}) in the previous Section~\ref{phase}. As discussed there, the LD-MC transition is a discontinuous one, and hence the long time averaged stationary density at the LD-MC coexistence is a DDW. Then by using the calculational framework developed above, the DDW height
 \begin{equation}
  \Delta (x) = \rho_\text{MC}(x) - \rho_\text{LD}(x),
 \end{equation}
and the probability $P(x)$ to find a DDW at $x$ is
\begin{equation}
 P(x)\propto [\rho_\text{MC}(x) - \rho_\text{LD}(x)]^2.
\end{equation}
In the present problem, $\rho_\text{MC}(x)=\rho_\text{HD}(x,J_\text{HD}=q_\text{min}/4)$. We now evaluate $\rho_\text{DDW}(x)$ for (\ref{qx-ddw1}). For $q(x)$ in (\ref{qx-ddw1}),
\begin{equation}
 \rho_\text{DDW}(x)=A\left(\frac{x}{2}-\frac{x^2}{4}\right)+B.\label{ddw-1}
\end{equation}
The constants $A$ and $B$ can be evaluated as follows. 
At $x=0$, i.e., the entry end of the TASEP lane, $\rho_\text{DDW}(x=0)=\alpha$, and $\rho_\text{DDW}(x=1)=1/2$. The last condition arises as with $q(x)=q_\text{min}$ at $x=1$ for (\ref{qx-ddw1}), both $\rho_\text{LD}(x=1)$ and $\rho_\text{HD}(x=1)$ and hence $\rho_\text{MC}(x=1)$ have the same value of 1/2. With these boundary conditions, we get
\begin{equation}
 B=\alpha,\, A=4\left(\frac{1}{2}-\alpha\right).
\end{equation}
In Fig.~\ref{ldmc-ddw1}(a), plots of $\rho_\text{DDW}(x)$ versus $x$ as obtained from our MFT and MCS studies are shown. To establish the existence of a delocalised domain wall at the LD-MC phase transition, we have also calculated the mean density $\bar\rho_s(t)$ {\em of the DDW} as a function of time $t$; here the subscript $s$ refers to spatial averages taken over the span of the DDW, same as the size of the TASEP lane in this case. When there is a DDW, $\bar\rho_s(t)$ fluctuates significantly, in contrast to its behaviour in the absence of any DDW. This is shown in Fig.~\ref{ldmc-ddw1}(b), which also includes the corresponding result for $\bar\rho_s(t)$ calculated close to the HD-MC transition (which according to our previous analysis is a continuous transition and hence the density is {\em not} a DDW), average done over the entire TASEP lane and shows very little fluctuations. Lastly, we also show a plot of $\Delta(x)$ versus $x$  at the LD-MC transition in Fig.~\ref{ldmc-ddw1}(c). This plot shows a large DDW height near $x=0$, progressively decreasing and ultimately vanishing at $x$ approaches 1.
\begin{figure}[htb]
 \includegraphics[width=\columnwidth]{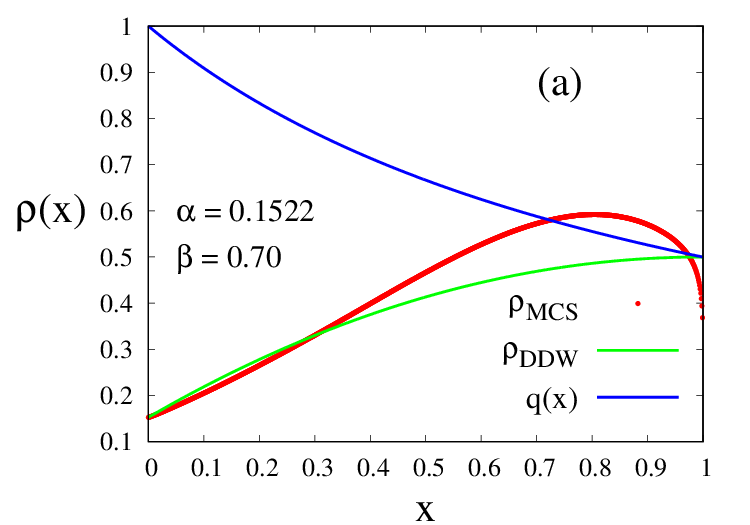}\\
 \includegraphics[width=0.50\columnwidth]{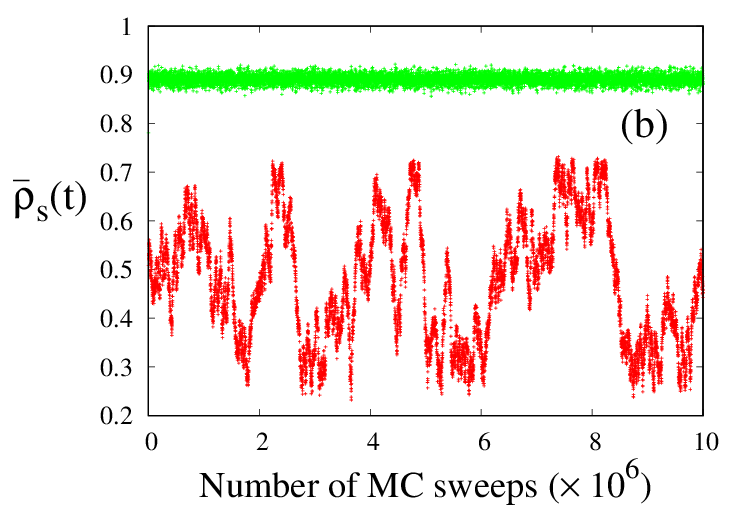}\hfill
 \includegraphics[width=0.48\columnwidth]{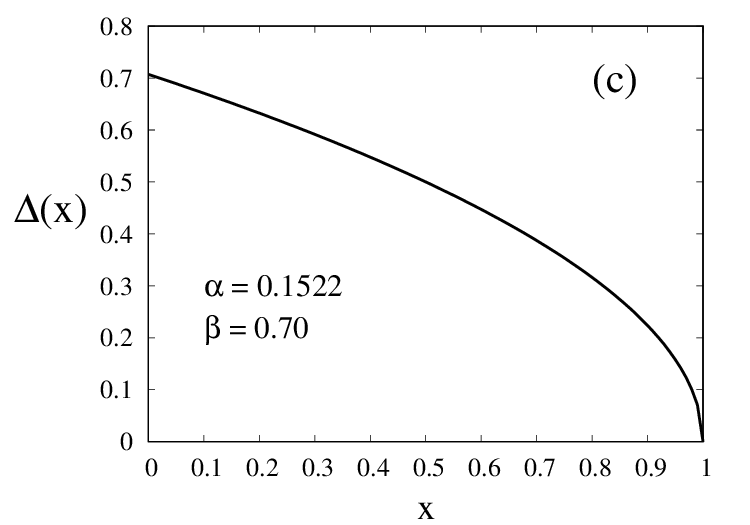}
 \caption{DDW at the LD-MC transition for $q(x)$ in (\ref{qx-ddw1}). (a) Plot of $\rho_\text{DDW}(x)$ versus $x$. (b) Plot of fluctuations in the $\bar\rho_s(t)$ versus time $t$ at the LD-MC transition (red, $\alpha=0.1522,\beta=0.70$), which strongly fluctuates in time. This is to be contrasted with $\bar\rho_s(t)$ versus time $t$ close to the HD-MC transition (green, $\alpha=0.70,\beta=0.1522$), which is significantly smoother, as expected in the absence of a DDW (see text). (c) Plot of $\Delta(x)$ versus $x$, which shows strong space dependence. } \label{ldmc-ddw1}
\end{figure}
The HD-MC transition for $q(x)$ in (\ref{qx-ddw1}) is a continuous transition. Hence, there is no DDW there.

(ii) $q(x)=\frac{1}{2-x}$, which is (\ref{qx-ddw2}) in the previous Section~\ref{phase}. In contrast to the previous case and as discussed in Section~\ref{phase}, the HD-MC transition is a discontinuous one, and hence the long time averaged stationary density at the HD-MC coexistence is a DDW. Then by using the calculational framework developed above, the DDW height for $q(x)$ in (\ref{qx-ddw2}) is given by
\begin{equation}
  \Delta (x) = \rho_\text{MC}(x) - \rho_\text{HD}(x,J=q_\text{min}),
 \end{equation}
and the probability $P(x)$ to find a DDW at $x$ is
\begin{equation}
 P(x)\propto [\rho_\text{MC}(x) - \rho_\text{HD}(x,J=q_\text{min}]^2.
\end{equation}
In the present problem, $\rho_\text{MC}(x)=\rho_\text{LD}(x,J_\text{LD}=q_\text{min}/4)$. We can now calculate $\rho_\text{DDW}(x)$ for $q(x)$ in (\ref{qx-ddw2}). We obtain
\begin{equation}
 \rho_\text{DDW}(x)=A\frac{x^2}{4} +B.\label{ddw-2}
\end{equation}
The boundary conditions on $\rho_\text{DDW}(x)$, which are to be used to find $A,B$, are as follows. At $x=1$, $\rho_\text{DDW}(x)=1-\beta$, since at $x=1$, $\rho_\text{MC}(x)=\rho_\text{HD}(x)=1-\beta$. At $x=0$, i.e., the entry end, $\rho_\text{DDW}(x)=1/2$ (ignoring the boundary layer, which has a vanishing thickness in the thermodynamic limit, there), since {\em both} $\rho_\text{LD}(x)$ and $\rho_\text{HD}(x)$ should be 1/2 there. Using these boundary conditions, we get 
\begin{eqnarray}
 &&B=1-\beta, A=-8\left(\beta - \frac{1}{2}\right).
\end{eqnarray}
In Fig.~\ref{hdmc-ddw2}(a) a plot of $\rho_\text{DDW}(x)$ versus $x$ for $q(x)$ in (\ref{qx-ddw2}) is given. Similar to the previous case and to establish a DDW at the HD-MC phase boundary, we have also calculated the mean density $\bar\rho_s(t)$ of the DDW at the HD-MC boundary as a function of time $t$ with the average being done over the span of the DDW, same as the size of the TASEP lane; it shows a strongly fluctuating behaviour, a telltale signature of a DDW.  This is shown in Fig.~\ref{hdmc-ddw2}(b). The plot in Fig.~\ref{hdmc-ddw2}(b) also includes the corresponding result for $\bar\rho_s(t)$ calculated close to the LD-MC transition, again averaged over the entire TASEP lane,  that shows very little fluctuations. This is consistent with  our
previous analysis that the LD-MC transition in the present case is a continuous transition and hence the density profile at the LD-MC transition is
not a DDW. Furthermore, we show a plot of the DDW height $\Delta(x)$ versus $x$ in Fig.~\ref{hdmc-ddw2}(c). This plot shows a large DDW height near $x=0$, progressively decreasing and ultimately vanishing at $x$ approaches 1. Notice that the DDW in Fig.~\ref{hdmc-ddw2}(a) is connected to the DDW in Fig.~\ref{ldmc-ddw1}(a) by the particle-hole symmetry,a reflection of the fact that $q(x)$ in (\ref{qx-ddw2}) is the particle-hole symmetric counterpart of $q(x)$ in (\ref{qx-ddw1}).


\begin{figure}[htb]
\includegraphics[width=\columnwidth]{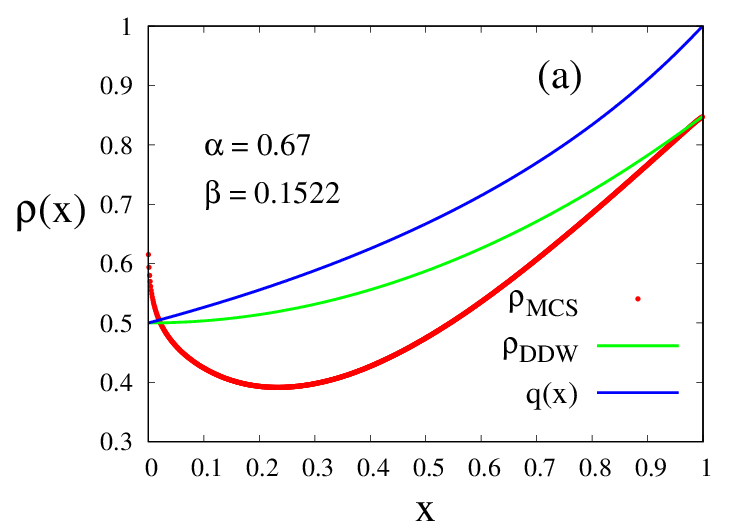}\\
\includegraphics[width=0.50\columnwidth]{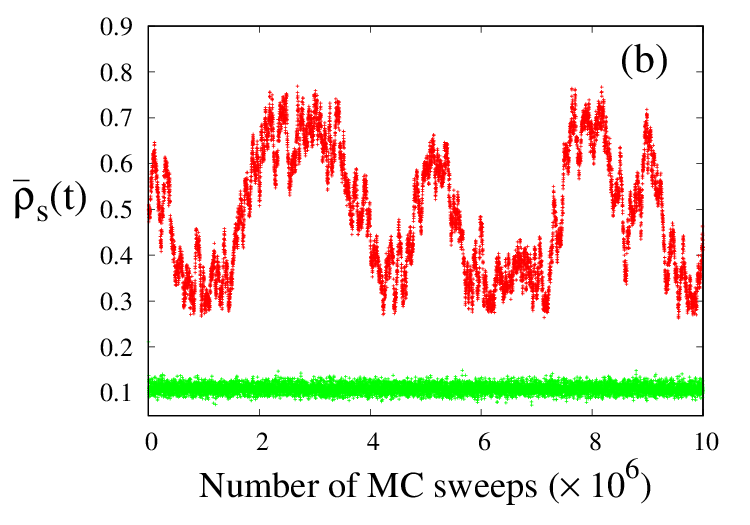}\hfill 
\includegraphics[width=0.48\columnwidth]{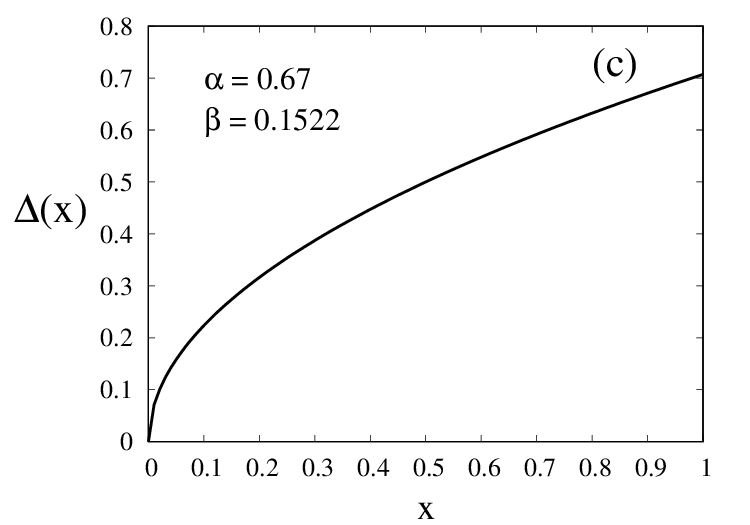}
 \caption{DDW at the HD-MC transition for $q(x)$ in (\ref{qx-ddw2}). (a) Plot of $\rho_\text{DDW}(x)$ versus $x$. (b)Plot of fluctuations in the $\bar\rho_s(t)$ versus time $t$ at the HD-MC transition (red, $\alpha=0.67, \beta=0.1522$), which strongly fluctuates in time. This is to be contrasted with $\bar\rho_s(t)$ versus time $t$ close to the LD-MC transition (green, $\alpha=0.1522,\beta=0.67$), which is significantly smoother, as expected in the absence of a DDW (see text). 
 (c) Plot of $\Delta(x)$ versus $x$. } \label{hdmc-ddw2}
\end{figure}

(iii) Finally, we consider
\begin{eqnarray}
 q(x)&=& \frac{1}{1+2x},\;\;\;0\leq x\leq 1/2,\nonumber \\
 &=& \frac{1}{3-2x},\;\;\;1/2\leq x\leq 1,
\end{eqnarray}
which is (\ref{qx-ddw3}) in Section~\ref{phase}. As explained in that Section, both the LD-MC and HD-MC transitions are discontinuous. The a DDW should be observed at both these transitions. The key difference is that now the DDWs at the LD-MC and HD-MC transitions are {\em partial}, i.e., covers on a part of the TASEP lane. We recall that for $q(x)$ in (\ref{qx-ddw3}), $\rho_\text{MC}(x)$ is given by (\ref{rho-mc-ddw3}).

We now calculate $\rho_\text{DDW}(x)$ in the present case. 

\subsubsection{Partial DDWs at the LD-MC transition}

Since $\rho_\text{MC}(x)=\rho_\text{HD}(x, J=q_\text{min}/4)$ for $0\leq x\leq 1/2$, there is a coexistence of $\rho_\text{LD}(x,J=q_\text{min}/4)$ and $\rho_\text{HD}(x,J=q_\text{min}/4)$ in that range of $x$ at the LD-MC transition. Then
following the analysis in Section~\ref{phase}, the DDW height at the LD-MC transition for $q(x)$ in (\ref{qx-ddw3}) is given by
\begin{equation}
 \Delta =  \rho_\text{MC}(x) - \rho_\text{LD}(x,J=q_\text{min}/4),
 \end{equation}
and the probability $P(x)$ to find a DDW at $x$ is
\begin{equation}
 P(x)\propto [\rho_\text{MC}(x) - \rho_\text{LD}(x,J=q_\text{min})]^2,
\end{equation}
for $0\leq x\leq 1/2$. For $1/2\leq x\leq 1$, $\rho_\text{MC}(x)=\rho_\text{LD}(x, J=q_\text{min}/4)$, giving $\Delta(x)=0,\,P(x)=0$ and hence no DDW in that range of $x$ at the LD-MC transition.
The DDW envelop at the LD-MC boundary has the form 
\begin{equation}
 \rho_\text{DDW}(x)=A\left(\frac{x}{2}-\frac{x^2}{2}\right) + B,\label{ddw-3a}
\end{equation}
for $x=0$ and $x=1/2$, since the DDW is confined only between $x=0$ and $x=1/2$. It is thus a {\em partial} DDW. We now need the boundary conditions at $x=0,1/2$. We have at $x=0$, $\rho_\text{DDW}(x)=\alpha$ and at $x=1/2$, $\rho_\text{DDW}(x)=1/2$. Applying these boundary conditions, we obtain 
\begin{equation}
 B=\alpha,\,A=8\left(\frac{1}{2}-\alpha\right).
\end{equation}

In Fig.~\ref{ldmc-ddw3a}(a) a plot of the long time averaged stationary density profile at the LD-MC transition as a function of $x$ for $q(x)$ in (\ref{qx-ddw3}) is given. This consists of a partial DDW with an envelop $\rho_\text{DDW}(x)$ in \eqref{ddw-3a} versus $x$  between $0\leq x\leq 1/2$ and $\rho_\text{MC}(x)=\rho_\text{LD}(x,J=q_\text{min}/4)$~\cite{atri-jstat} for $1/2\leq x\leq 1$.  
To establish a partial DDW at the LD-MC phase boundary, we have also calculated the mean density $\bar\rho_s(t)$ of the DDW, averaged over the range $0\leq x\leq 1/2$, at the LD-MC boundary as a function of time $t$, which shows a strongly fluctuating behaviour, a telltale signature of a DDW. This is shown in Fig.~\ref{ldmc-ddw3a}(b), marked in strongly fluctuating red lines. A plot of $\bar\rho_s(t)$ calculated in the segment $1/2\leq x\leq 1$ is also shown in Fig.~\ref{ldmc-ddw3a}(b), which shows very little fluctuations, as there is no DDW between $x=1/2$ and $x=1$ (green line).   This is consistent with  our
above analysis that there is a {\em partial} DDW at the LD-MC transition only for $0\leq x\leq 1/2$. Furthermore, we show a plot of the DDW height $\Delta(x)$ versus $x$ in Fig.~\ref{ldmc-ddw3a}(c). This plot shows a large DDW height near $x=0$, which decreases, ultimately vanishing at $x$ approaches 1/2. Between 1/2 and 1, $\Delta(x)=0$, which is consistent with the fact that the DDW is confined only between $x=0$ and $x=1/2$.


\begin{figure}[htb]
 \includegraphics[width=\columnwidth]{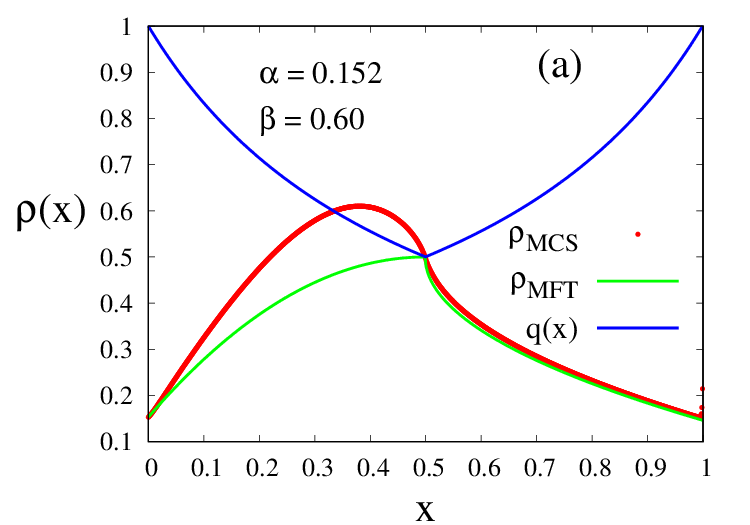}\\
 \includegraphics[width=0.48\columnwidth]{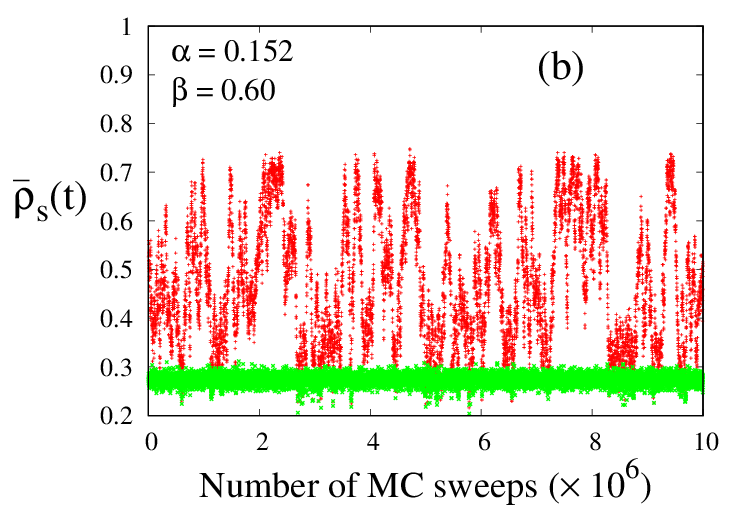}\hfill
 \includegraphics[width=0.48\columnwidth]{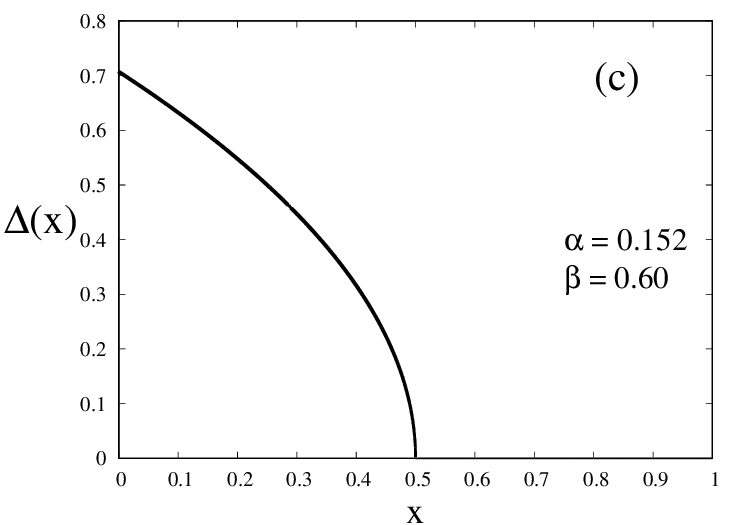}
 \caption{Partial DDW at the LD-MC transition for $q(x)$ in (\ref{qx-ddw3}).  (a) Plot of $\rho_\text{DDW}(x)$ versus $x$. It extends only from $x=0$ to $x=1/2$. (b) Plot of $\bar\rho_s(t)$ versus time $t$ in the segment $0\leq x\leq 1/2$ (red), which shows strong fluctuates in time.  Plot of $\bar\rho_s(t)$ versus time $t$ in the segment $1/2\leq x\leq 1$ is also shown for comparison (green), which is significantly smoother. (d) Plot of $\Delta(x)$ versus $x$, which is nonzero only between $x=0$ to $x=1/2$ for $q(x)$ in (\ref{qx-ddw3}); $\Delta(x)=0$ for $1/2\leq x\leq 1$ (see text).}\label{ldmc-ddw3a}
\end{figure}


\subsubsection{Partial DDWs at the HD-MC transition}

We now consider the envelop of a partial DDW formed at the HD-MC phase transition. In this case, since $\rho_\text{MC}(x)=\rho_\text{LD}(x, J=q_\text{min}/4)$ for $1/2\leq x\leq 1$, there is a coexistence of $\rho_\text{LD}(x,J=q_\text{min}/4)$ and $\rho_\text{HD}(x,J=q_\text{min}/4)$ in the above range of $x$ at the HD-MC transition. Then
following the analysis in Section~\ref{phase}, the DDW height at the HD-MC transition for $q(x)$ in (\ref{qx-ddw3}) is given by
\begin{equation}
 \Delta =  \rho_\text{HD}(x,J=q_\text{min}) - \rho_\text{MC}(x),
 \end{equation}
and the probability $P(x)$ to find a DDW at $x$ is
\begin{equation}
 P(x)\propto [\rho_\text{HD}(x,J=q_\text{min}) - \rho_\text{MC}(x)]^2,
\end{equation}
for $1/2\leq x\leq 1$. For $0\leq x\leq 1/2$, $\rho_\text{MC}(x)=\rho_\text{HD}(x, J=q_\text{min}/4)$, giving $\Delta(x)=0,\,P(x)=0$ and hence no DDW in that range of $x$ at the HD-MC transition.
The DDW envelop at the HD-MC boundary has the form 
\begin{equation}
 \rho_\text{DDW}(x)=A\left(-\frac{x}{2}+\frac{x^2}{2}\right) + B,\label{ddw-3b}
\end{equation}
for $x=1/2$ and $x=1$, since the DDW is confined only between $x=1/2$ and $x=1$. It is thus a {\em partial} DDW. We now need the boundary conditions at $x=0,1/2$. We have at $x=1$, $\rho_\text{DDW}(x)=1-\beta$ and at $x=1/2$, $\rho_\text{DDW}(x)=1/2$. Applying these boundary conditions, we obtain 
\begin{equation}
 B=1-\beta,\,A=8\left(\frac{1}{2}-\beta\right).
\end{equation}

In Fig.~\ref{hdmc-ddw3b}(a), we give a plot of the long time averaged stationary density profile at the HD-MC transition as a function of $x$ for $q(x)$ in (\ref{qx-ddw3}). This consists of $\rho_\text{MC}(x)=\rho_\text{HD}(x,J=q_\text{min}/4)$~\cite{atri-jstat} for $0\leq x\leq 1/2$ and a partial DDW with an envelop $\rho_\text{DDW}(x)$ in \eqref{ddw-3b} versus $x$  between $1/2\leq x\leq 1$.  
To establish a partial DDW at the HD-MC phase boundary, we have also calculated the mean density $\bar\rho_s(t)$ of the DDW, averaged over the range $1/2\leq x\leq 1$, at the HD-MC boundary as a function of time $t$, which shows a strongly fluctuating behaviour, a telltale signature of a DDW. We present the plot in Fig.~\ref{hdmc-ddw3b}(b) (red), which is strongly fluctuating in time. This is to be contrasted  with $\bar\rho_s(t)$, averaged over the segment $0\leq x\leq 1/2$, which shows very little fluctuations, as there is no DDW between $x=0$ and $x=1/2$; see Fig.~\ref{hdmc-ddw3b}(b) (green).   This is consistent with  our
above analysis that there is a {\em partial} DDW at the HD-MC transition for $1/2\leq x\leq 1$. Furthermore, we show a plot of the DDW height $\Delta(x)$ versus $x$ in Fig.~\ref{hdmc-ddw3b}(c). This plot shows a vanishingly small DDW height near $x=1/2$, which increases monotonically as $x$ approaches 1. Between 0 and 1/2, $\Delta(x)=0$, which is consistent with the fact that the DDW is confined only between $x=1/2$ and $x=1$. Notice that the symmetry of $q(x)$ in (\ref{qx-ddw3}) about $x=1/2$ is reflected in the complementary natures of the DDWs in Fig.~\ref{ldmc-ddw3a}(a) and Fig.~\ref{hdmc-ddw3b}(a).

\begin{figure}[htb]
 \includegraphics[width=\columnwidth]{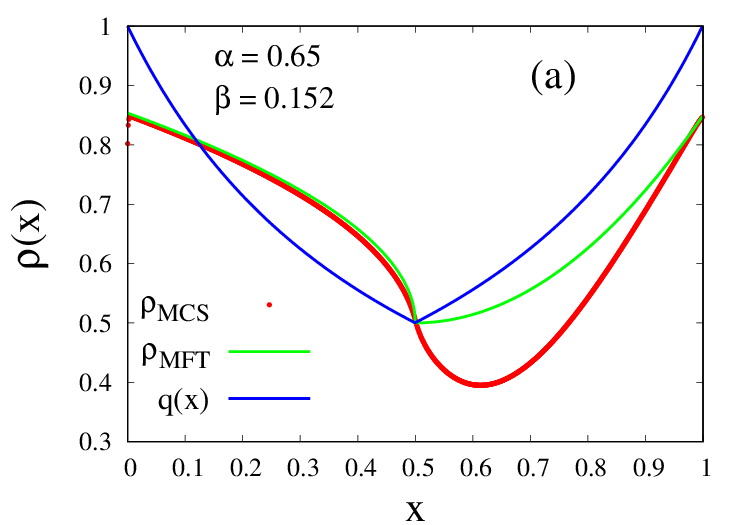}\\
 \includegraphics[width=0.48\columnwidth]{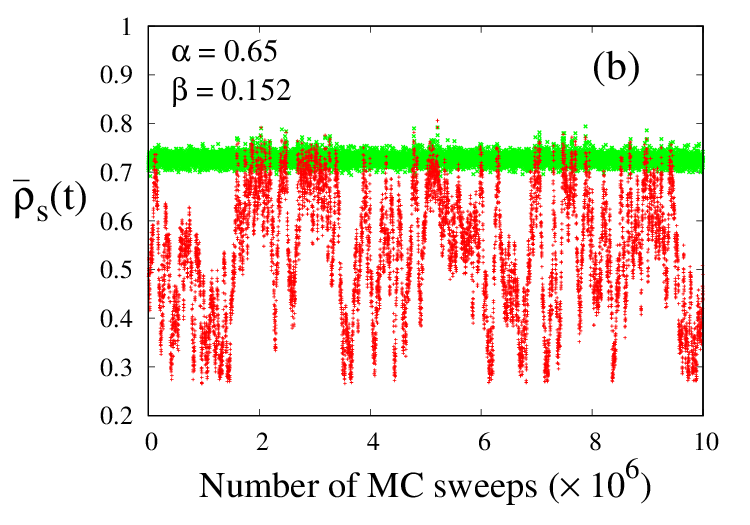}\hfill
 \includegraphics[width=0.48\columnwidth]{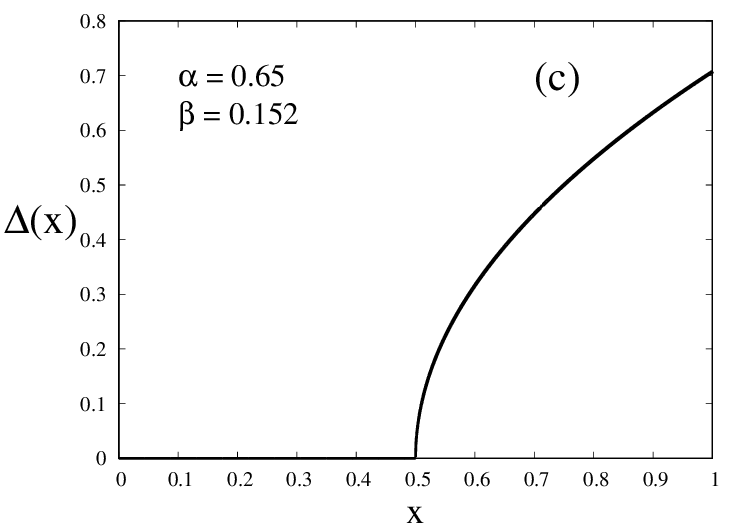}
 \caption{Partial DDW at the HD-MC transition for $q(x)$ in (\ref{qx-ddw3}).  (a) Plot of $\rho_\text{DDW}(x)$ versus $x$. It extends only from $x=1/2$ to $x=1$. (b) Plot of $\bar\rho_s(t)$ (red), averaged over $0\leq x\leq 1/2$, versus time $t$, which fluctuates strongly in time.  The plot of $\bar\rho_s(t)$ (green), averaged over $1/2\leq x\leq 1$, versus time $t$  is also shown (green), which is smooth. (c) Plot of $\Delta(x)$ versus $x$ between $x=1/2$ to $x=1$; $\Delta(x)=0$ for $0\leq x\leq 1/2$. See text.} 
 \label{hdmc-ddw3b}
\end{figure}


\subsection{Partial DDWs at the triple point}

Lastly, we consider the DDWs at the triple point, which is the meeting point of the LD-HD, LD-MC and HD-MC phase boundaries. We focus on the case when all the three are discontinuous transitions. We shall see that the long time averaged density profile at the triple point has an exotic form. As a concrete example of this case, we use $q(x)$ as given in (\ref{qx-ddw3}) for which {\em all} the three transitions are discontinuous (see above). In particular, $\rho_\text{MC}(x)$ is given by (\ref{rho-mc-ddw3}). Therefore, there should be density coexistence, i.e., coexistence of $\rho_\text{LD}(x,J=q_\text{min}/4)$ and $\rho_\text{LD}(x,J=q_\text{min}/4)$ along the {\em entire} TASEP lane. This means the long time averaged stationary density profile along the whole of the TASEP lane is a DDW that we show has some unusual features. 

Following the calculational set up in the previous Section, we find that the DDW height $\Delta (x)$ is given by
\begin{equation}
 \Delta =  \rho_\text{HD}(x,J=q_\text{min}) - \rho_\text{MC}(x),
 \end{equation}
and the probability $P(x)$ to find a DDW at $x$ is
\begin{equation}
 P(x)\propto [\rho_\text{HD}(x,J=q_\text{min}) - \rho_\text{MC}(x)]^2,
\end{equation}
 along the entire TASEP lane ($0\leq x\leq 1$). Since for $q(x)$ in (\ref{qx-ddw3}), $\rho_\text{LD}(x=1/2)=1/2=\rho_\text{HD}(x=1/2)$, we have {\em both} height $\Delta(x)$ and probability $P(x)$ for the DDW vanishing as $x\rightarrow 1/2$. Effectively, this means there are {\em two} DDWs, one in the range $0\leq x\leq 1/2$, arising out of the phase coexistence between $\rho_\text{MC}(x)$ and $\rho_\text{LD}(x)$, and the second one in the range $1/2\leq x\leq 1$, arising out of the coexistence between $\rho_\text{HD}(x)$ and $\rho_\text{MC}(x)$. The two DDWs are further distinguished by their {\em different} time-averaged mean densities, $\bar \rho_L(t)$ in the left half that is averaged over $x=0$ and $x=1/2$, and $\bar \rho_R(t)$ in the right half that is averaged over $x=1/2$ and $x=1$, respectively. The boundary conditions are (i) $\rho_\text{DDW}^L(x=0)=\alpha$, (ii) $\rho_\text{DDW}^L(x=1/2)=1/2$, (iii) $\rho_\text{DDW}^R(x=1/2)=1/2$, (iv) $\rho_\text{DDW}^R(x=1)=1-\beta$, where $\rho_\text{DDW}^L(x)$ and $\rho_\text{DDW}^R(x)$ are the DDW envelops, respectively, for $0\leq x\leq 1/2$ and $1/2\leq x\leq 1$.  Following the calculational set up discussed above, we find
 \begin{eqnarray}
  &&\rho_\text{DDW}^L(x)= 4 \left(\frac{1}{2}-\alpha\right)(x-x^2)+\alpha,\,0\leq x\leq 1/2\\
  &&\rho_\text{DDW}^R(x)= 4 \left(\frac{1}{2}-\beta\right)(-x+x^2)+1-\beta,\,1/2\leq x\leq 1.
 \end{eqnarray}

 In Fig.~\ref{triple-ddw}(a) a plot of the DDWs at the triple point as a function of $0\leq x\leq 1$ for $q(x)$ in (\ref{qx-ddw3}) is given.   To establish the delocalised nature of the domain walls at the triple point, we have also calculated the mean densities $\bar\rho_L(t)$  and $\bar\rho_R(t)$ of the DDWs in the left and right halves of the TASEP lane, averaged over  $0\leq x\leq 1/2$ and $1/2\leq x\leq 1$ for the left and right halves respectively as functions of time $t$, both of which show a strongly fluctuating behaviour, a telltale signature of a DDW. These are shown in Fig.~\ref{triple-ddw}(b) and Fig.~\ref{triple-ddw}(c) respectively.  We have also shown the respective time-averaged mean densities in Fig.~\ref{triple-ddw}(b) and Fig.~\ref{triple-ddw}(c). The two mean densities have distinct values, which distinguish the DDWs in the two halves. In Fig.~\ref{height-triple-ddw}, we have shown a plot of the domain wall height versus $x$, which vanishes at $x=1/2$, and reaches its maximum at $x=0, 1$. This feature supports the idea of two (partial) DDWs connected at $x=1/2$.

 \begin{figure}[htb]
  \includegraphics[width=\columnwidth]{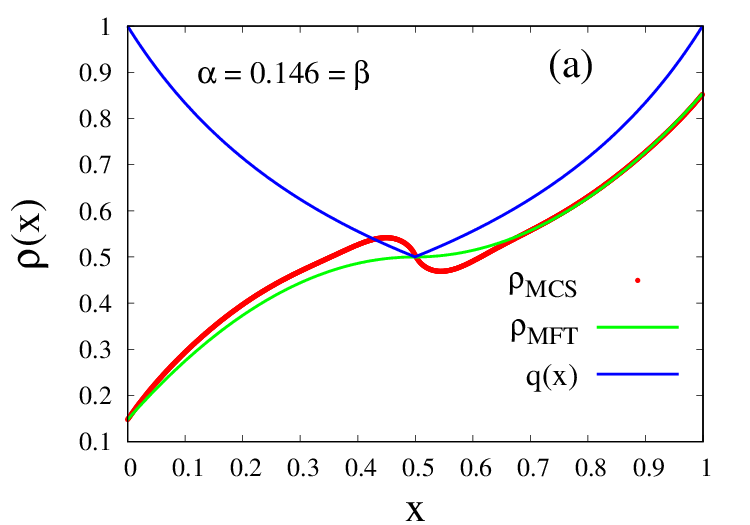}\\
  \includegraphics[width=0.48\columnwidth]{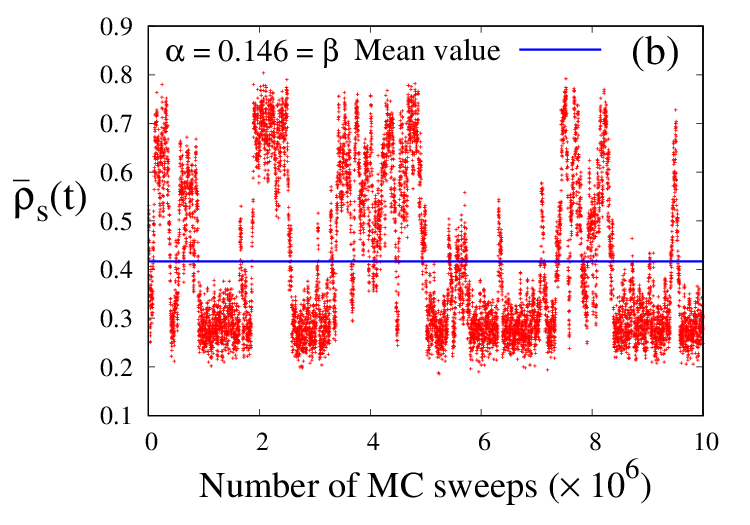}\hfill
  \includegraphics[width=0.48\columnwidth]{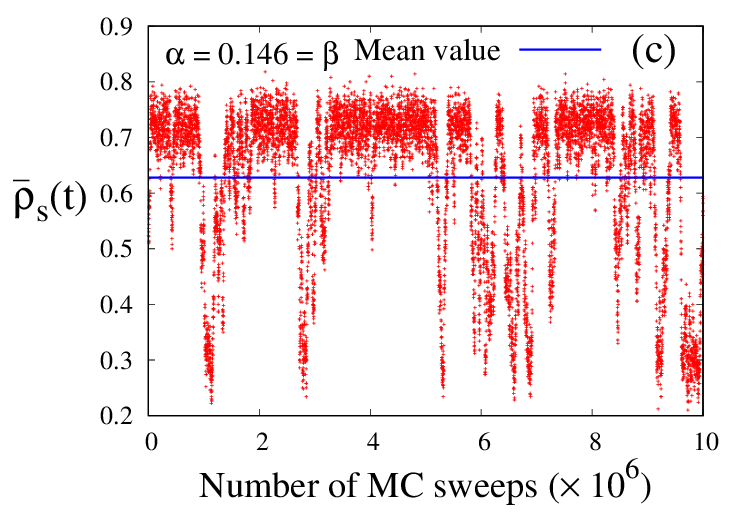}
  \caption{DDWs at the triple point. (a) Plots of the long-time averaged DDW envelops as functions of $x$. (b) Plots of   $\bar\rho_L(t)$ versus time $t$ in the segment $0\leq x\leq 1/2$. Time-averaged $\bar\rho_L(t)$, a constant, is marked by a straight line (blue). (c) Plot of $\bar\rho_L(t)$ versus time $t$ in the segment $1/2\leq x\leq 1$. Time-averaged $\bar\rho_R(t)$, another constant, is marked by a straight line (blue). See text. }\label{triple-ddw}
 \end{figure}
  
 \begin{figure}[htb]
  \includegraphics[width=\columnwidth]{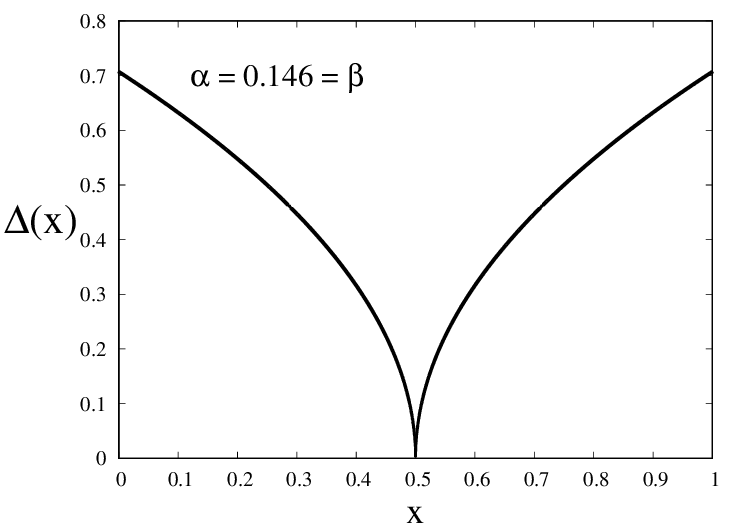}
  \caption{Plot of $\Delta(x)$ versus $x$, which vanishes at $x=1/2$ and reaches its maximum at the extreme ends $x=0,1$. See text.}\label{height-triple-ddw}
 \end{figure}

}
 
 \section{Monte-Carlo simulations}\label{monte}

Our model as defined in Section~\ref{model} above consists of  a lattice of $L$ sites, each of which is labelled by an index $i$ with $i\in [1,L]$.
 We represent the occupation at site $i$ at time $t$ by $n_i(t)$, which is either 0 or 1.  We perform MCS studies of the model subject to the update rules (a)-(c) described above in Sec.~\ref{model} by using a random sequential updating scheme. The particles enter the system through the left most site ($i=1$) at a fixed rate $\alpha q(1)$, where $\alpha$ a control parameter of the model, if it is vacant. Then after hopping through the system from $i=1$ to $L$, subject to exclusion, the particles exit the system from $i=L$ at a fixed rate $\beta q(L)$, where $\beta$ is another control parameter of the model. Here, $\alpha$ and $\beta$  are varied to produce different steady states. After reaching the steady states, the density profiles are calculated and temporal averages are performed. This produces time-averaged, space-dependent density profiles, given by $\langle n_i\rangle$,  parametrised by $\alpha$ and $\beta$; here $\langle...\rangle$ implies temporal averages over the steady states of the model. We have performed with  $L=10000$ ($L=1000$ for the domain walls) up to $10^7$ Monte-Carlo steps. 
 { The steady states are reached by the system after spending certain transient times. We make a technically important point here. In an open uniform TASEP, the existence of a steady state can be easily confirmed by visually observing the spatio-temporal constancy of the average density $\langle n_i(t)\rangle$ (excluding the domain walls) in the bulk of the system. In the present problem, this is no longer true due to the  spatially varying steady state density in the bulk. Instead, we study the spatial constancy of the current $J$ in the steady state, a condition that holds both in the present study and also for a uniform open TASEP. In our MCS studies, we make all our measurements  only after this condition is satisfied, i.e., a constant $J$ is achieved. This automatically ensures that all the numerical measurements reported above correspond to steady state quantities.} 

\section{Summary and conclusions}\label{summ}

To summarise, we have studied the stationary densities and domain walls in open inhomogeneous TASEPs, characterised by hopping rates, which vary slowly in space, but can have isolated discontinuities. Building upon and extending the results discussed in Ref.~\cite{atri-jstat}, we have obtained the stationary densities, which are generically space-dependent, and depend crucially on the precise space dependence of the hopping rate functions $q(x)$. For the particular choices  of $q(x)$ made in the present study, we have shown that the steady state MC phase density $\rho_\text{MC}(x)$ can be {\em discontinuous}, even when $q(x)$ is {\em continuous}. We have shown how MFT arguments can be made in these cases, which are corroborated by our MCS simulations. 

We have constructed the mean-field phase diagrams in the $\alpha-\beta$ plane, which reveal a remarkable degree of universality in their topology, independent of the forms of $q(x)$. In fact, in one case, the phase diagram matches {\em exactly} with that for an open, uniform TASEP. However, all the phase boundaries in this case represent discontinuous transitions. Remarkably, even when there is a continuous transition, the mean-field order parameter exponent is found to be different from its open uniform TASEP counterpart. This shows the inadequacy of the phase diagrams as markers or proxies for $q(x)$. 
Even the knowledge of the phase transitions may not suffice. For instance, one can construct another $q(x)$ that has the same symmetry as in the case I in the Section~\ref{phase}, giving a phase diagram identical to that in an open, uniform TASEP and having all discontinuous transitions as in the case I. Thus, one needs to {\em actually} measure the full space-dependent stationary densities to learn about the underlying $q(x)$. This is an important outcome from the present study. 

We have studied the delocalised domain walls (DDW) in details. Unlike in an open TASEP with uniform hopping rates, for which DDWs can be found only at the LD-HD phase transition, in the present problem, DDWs can be found even at the LD-MC or HD-MC phase transitions. This is because any (or both) of these two transitions can be discontinuous in nature, depending on the form of $q(x)$. A particularly intriguing situation can arise when $q(x)$ have a minimum in the bulk and consequently {\em both} of LD-MC and HD-MC transitions are discontinuous. In this case, the resulting DDWs at the LD-MC and HD-MC transitions are {\em partial}, i.e., they extend over only a segment of the TASEP lane.  We have calculated the DDW envelopes at all of these transitions analytically, which are generally nonlinear functions of $x$. Studying this particular case further, we investigated the long time stationary density profiles at the triple point, which is the meeting point of all the three discontinuous transition for this particular choice of $q(x)$. Due to the discontinuous nature of all the three transitions, there is phase coexistence of the distinct steady state density solutions at this point in the phase diagram. Although there are three distinct phases possible, there are actually only {\em two} distinct solutions for the steady state densities at the triple point, which are $\rho_\text{LD}(x)$ and $\rho_\text{HD}(x)$; $\rho_\text{MC}(x)$ is {\em not} an independent solution, but is constructed out of the other two.  We have shown that these considerations lead to the long time averaged density profile at the triple point to be effectively viewed two DDWs connected serially at $x=x_0=1/2$, which is the location of the minimum of $q(x)$ and where the probability to find each of the DDW vanishes. While our explicit calculations were done for a simple choice with $x_0=1/2$, our calculational framework indicates that the qualitative features of these results hold independent of where $x_0$ is located in the bulk. In general, the different choices for $q(x)$ used here allow us to connect their analytical properties with those of the DDW envelops through our analytical predictions, corroborated by the MCS results. We notice a greater mismatch between the MFT and MCS results on the DDWs at LD-MC and HD-MC transitions than at the LD-HD transition. We attribute this feature to the usually more pronounced boundary layers for $\rho_\text{MC}(x)$, which are always ignored in the MFT.
Application of boundary layer theories, originally developed in Refs.~\cite{boun1,boun2,boun3} could be useful in this context and provides a promising future direction.


{  Our studies above are confined to considering a few simple choices for $q(x)$ with specific analytic structures. These can be extended to more complex forms for $q(x)$. For instance, even with $q(x)$ having additional maxima or minima, our analytical calculations on the phase diagrams already suggest that the phase diagrams with more complex forms for $q(x)$ are still determined by $J_\text{in}$ and $J_\text{out}$, the injection limited and extraction limited stationary currents, along with the same MFT scaling exponents in case of continuous transitions. Thus all possible $q(x)$ with the same $J_\text{in}$ and $J_\text{out}$ should give {\em exactly} the same phase diagram, reinforcing the idea of universal phase diagrams; see also Ref.~\cite{atri-jstat}. Secondly, for $q(x)$ with multiple discontinuities should give rise to DDW envelops with those many kinks, a prediction testable in MCS studies. }

Our studies here can be extended in a various directions. It will be interesting to investigate the stationary densities and domain walls, when $q(x)$ has multiple global minima. One may also consider one or more point or extended defects~\cite{lebo,mustansir,niladri-tasep,tirtha-niladri,atri3,sourav4} within an otherwise smoothly varying hopping rate function $q(x)$, which can compete with the global minima of $q(x)$. The interplay between the two should control the resulting stationary densities. It will be interesting to extend our study to TASEPs with finite resources~\cite{reser1,reser2,astik-parna,sourav1,sourav2,AB-EF,sourav3} and see how smoothly varying hopping rates can affect and compete with finite resources. Yet another interesting extension would be to study the effects of interactions within our set up~\cite{akgupta2}. In the present work, we have only concerned ourselves with studying the static or stationary properties. It will be interesting to explore how time-dependent properties can depend upon various $q(x)$. 
We hope our work here will provide impetus to future studies along these directions.

\section{Acknowledgement}

S.M. and A.B. thank  Alexander von Humboldt Stiftung (Germany) for partial financial support  through their  Research Group Linkage programme (2024). S.M. thanks ANRF (India) for partial financial
support through the ARG (MATRICS) programme (file no.: ANRF/ARGM/2025/000748/TS).  A.B. thanks ANRF (India) for partial financial
support through the ARG (MATRICS) programme (file no.:
ANRF/ARGM/2025/000461/TS).





\end{document}